\def\CleanCopy{1}
\documentclass[
 aip,
 pof,
 amsmath,amssymb,
 preprint,
]{revtex4-1}

\usepackage{graphicx}
\usepackage{dcolumn}
\usepackage{bm}
\usepackage[mathlines]{lineno}

\usepackage[utf8]{inputenc}
\usepackage[T1]{fontenc}
\usepackage{mathptmx}
\usepackage{etoolbox}
\usepackage{xcolor}
\usepackage{comment}
\usepackage{subfig}
\usepackage[normalem]{ulem}
\usepackage{booktabs}
\graphicspath{{./figures/}}

\definecolor{REVBLUE}{RGB}{0,112,192}

\ifdefined\CleanCopy
  \newcommand{\NREV}[1]{#1}
\else
  \newcommand{\NREV}[1]{{\color{REVBLUE}#1}}
\fi
\ifdefined\CleanCopy
  \newcommand{\REV}[1]{#1}
  \newcommand{\REVCUT}[1]{}
  \newcommand{\MREV}[1]{#1}
\else
  \newcommand{\REV}[1]{{\color{REVBLUE}#1}}
  \newcommand{\REVCUT}[1]{{\color{REVBLUE}\sout{#1}}}
  \newcommand{\MREV}[1]{{\color{REVBLUE}#1}}
\fi
\newcommand{\ednote}[1]{}

\newcommand{\LANG}[1]{\REV{#1}}
\newcommand{\LANGCUT}[1]{\REVCUT{#1}}

\newcommand{\NEW}[1]{\REV{#1}}

\newcommand{\CUT}[1]{\REVCUT{#1}}
\newcommand{\FIX}[1]{\REV{#1}}
\newcommand{\FIXCUT}[1]{\REVCUT{#1}}

\newcommand{\rev}[1]{\REV{#1}}
\newcommand{\ed}[1]{\REV{#1}}
\newcommand{\rw}[1]{\REV{#1}}

\newcommand{\gr}[1]{\REV{#1}}
\newcommand{\tb}[1]{\REV{#1}}

\newcommand{\Rey}{\mathit{Re}}

\newcommand{\We}{\mathit{We}}
\newcommand{\Oh}{\mathit{Oh}}
\newcommand{\De}{\mathit{De}}
\newcommand{\ts}{t^{*}}
\newcommand{\Deq}{D_{eq}}
\newcommand{\Dh}{D_{h}}
\newcommand{\Db}{D_{b}}

\makeatletter
\def\@email#1#2{
 \endgroup
 \patchcmd{\titleblock@produce}
  {\frontmatter@RRAPformat}
  {\frontmatter@RRAPformat{\produce@RRAP{}\produce@RRAP{*#1\href{mailto:#2}{#2}}}\frontmatter@RRAPformat}
  {}{}
}
\makeatother

\begin{document}

\preprint{}

\title[]{\NREV{Experimental study of the impact dynamics of polymeric hollow droplets}}

\author{Mohammad Mahdi Nasiri}
\thanks{}
\affiliation{Department of Mechanical, Industrial \& Aerospace Engineering, Concordia University, 1455 De Maisonneuve Blvd. W., Montreal, Quebec, Canada}

\author{Mohammad Reza Daneshvar Garmroodi}
\thanks{}
\affiliation{Department of Civil Engineering, McMaster University, Hamilton, Ontario, Canada}

\author{Damian Vadillo}
\affiliation{3M Corporate Research Analytical Laboratory, Polymer Science Group, St.\ Paul, Minnesota, United States}

\author{Moussa Tembely}
 \email{moussa.tembely@concordia.ca}
\affiliation{Department of Mechanical, Industrial \& Aerospace Engineering, Concordia University, 1455 De Maisonneuve Blvd. W., Montreal, Quebec, Canada}

\begin{abstract}
\rev{The impact dynamics of hollow droplets, while influential in applications \FIXCUT{like} \FIX{such as} coating and spraying, remain less explored than their dense counterparts. \FIXCUT{particularly for non-Newtonian fluids.} \FIX{In particular, the impact dynamics of viscoelastic hollow droplets have yet to be fully explored.} This study presents an experimental investigation into the impact of hollow Newtonian (water) and viscoelastic (polymeric solution) droplets on a solid surface \NREV{at different impact velocities and polymer concentrations}. We demonstrate two hallmark features of hollow droplet flattening: the formation of a central counter-jet and the final deposition, \MREV{both associated with the entrapped air bubble}. \NREV{For Newtonian droplets}, the counter-jet exhibits rapid growth and breakup due to capillary instabilities. Introducing polymer additives fundamentally alters this behavior: viscoelasticity affects the counter-jet's height and velocity, delays bubble rupture, and inhibits droplet detachment. Crucially, we observe the emergence of beads-on-a-string structures during filament thinning, a signature of the competition between elastic and capillary forces. \rw{By systematically varying the polymer concentration and impact velocity, we identify the conditions under which three distinct outcomes occur: deposition, partial deposition, and \FIXCUT{rebound} \FIX{detachment}. Our results show how inertia, viscosity, capillarity, and elasticity together govern the splashing morphology of hollow non-Newtonian droplets.}}
\end{abstract}


\maketitle

\section{Introduction}
\label{sec:intro}

Droplet impingement on a solid or liquid surface is a fundamental phenomenon \NREV{occurring in a wide variety of applications}, including combustion, coatings deposition, inkjet printing, soil erosion, and air entrapment at the sea's surface \cite{josserand2016drop,thoroddsen2008high,liang2016review,zhu2020impact,zheng2021heat,gilet2012droplets,guo2020oblique,guan2021post,gordillo2019theory}. \NREV{Experimental, analytical, and numerical studies of droplet impact consider parameters such as impact velocity, size, and rheology} \cite{pasandideh1996capillary,liu2024simulation,isukwem2024role,shah2024drop,zhang2021effect}.

In general, both viscous dissipation and surface tension forces due to the deformation of the droplet and its interaction with the substrate determine the spreading ratio. \NREV{Upon impact, part of the kinetic energy of the droplet is transformed into surface energy} as the droplet expands to attain a maximal spreading diameter, corresponding to a pancake shape.

In some cases, the spreading drop's rim becomes unstable, \NREV{causing a secondary drop to be ejected, a process called splashing}. Following the initial spreading phase, a droplet may either preserve its radially extended configuration or undergo partial retraction. On hydrophobic substrates, particularly at high impact velocities, droplets can completely withdraw and in some cases even detach from the surface. Considerable research has been devoted to elucidating the mechanisms governing spreading \cite{laan2014maximum,lee2016universal,ukiwe2005maximum}, receding \cite{bartolo2005retraction,wang2020retraction}, and bouncing \cite{richard2002contact,riboux2014experiments}, as well as the complex transition from spreading to splashing \cite{josserand2016drop}.

Polymeric additives are used in many industrial processes to adjust fluid properties, \NREV{such as adjusting viscosity}, and to manage interactions between fluids and solid surfaces. Since polymers are common in natural systems, using them as fluid additives is both practical and effective. In industries like coating, spraying, and pesticide application, reducing droplet bounce and rebound is crucial for product efficiency. \NREV{The impact of polymeric droplets (liquid droplets containing a specified polymer concentration) on solid surfaces has received significant attention because of its diverse applications} such as lab-on-a-chip (LOC) \cite{asghari2020non}, criminology \cite{yokoyama2022droplet}, coating and spraying \cite{blossey2003self}. \NREV{Recent studies have also considered the impact of complex-fluid droplets} \cite{quirke2024spreading,tang2025droplet,han2024motion,biroun2023impact,mobaseri2025maximum}. Adding polymers is an effective way to change and control \NREV{how droplets behave after impacting surfaces}. For instance, the study by Bergeron \textit{et al.}\ showed that \NREV{even a small amount of polymer additive} can effectively suppress bounce of liquid droplets impacting a hydrophobic surface \cite{bergeron2000controlling}.

\MREV{At sufficiently low concentrations, polymer additives can have a modest effect on shear viscosity; however,} under extensional flow conditions the polymer chains stretch and deform, resulting in a pronounced increase in elongational viscosity as shown in filament thinning studies \cite{dinic2019macromolecular}.  \MREV{Because elongational stresses can affect the spreading and retraction of impacting droplets,} it has been debated whether polymer concentration should significantly influence droplet impact behavior. Experimental findings, however, indicate that polymer additives exert only a minor influence on the spreading phase \cite{an2012maximum}, while the retraction phase proceeds at a noticeably reduced velocity \cite{huh2015role}.

In addition to normal drops, there are compound drops that are \NREV{a relatively recent topic with many industrial applications} \cite{bertola2012single}, such as water-oil emulsions for steel strip manufacturing \cite{prunetfoch1998impacting} or polymer solutions \FIXCUT{for agricultural sprays} \FIX{in spray processes} \cite{lopezherrera2019adaptive} or additive manufacturing processes \cite{visser2018inair}.

While the dense droplet impingement has been the focus of many studies \cite{quirke2024spreading,tang2025droplet,han2024motion,biroun2023impact,mobaseri2025maximum,hao2015superhydrophobic,garciageijo2020inclined,quintero2019splashing}, only a minority have considered a hollow droplet impact on a surface \cite{gulyaev2013hollow,wei2021maximum,liu2021numerical,nasiri2020investigation,nasiri2024experimental,nasiri2021hollow,nasiri2023flattening} \REV{\cite{qian2025hollow,hu2025formation,zhu2024analysis,yan2025dynamic}}. Hollow droplets are often formed during the thermal spraying and diesel injection nozzles \cite{li2019numerical}. The impact of a hollow droplet on a solid surface is \NREV{fundamentally different from that of a dense droplet}. Therefore, the effects of \REV{hollow} droplets were studied in recent years. The entrapped bubble plays a significant role in the droplet's deformation behavior. During the early stages of impact, the bubble undergoes compression and subsequent collapse under the high pressure exerted by the droplet. The bursting of the bubble produces waves and disturbances on the liquid sheet, disrupting its stability. This disturbance often causes the sheet to rupture at its center, leading to the formation of a hole within the spreading film. Consequently, when the liquid begins to recoil toward the center, it fails to reform into a hemispherical shape. Instead, it stabilizes into a donut-like configuration on the surface. The generation of a counter-jet, bubble breakup, and the associated surface perturbations are characteristic behaviors of hollow droplet impacts and are consistently observed across surfaces with varying wettability \cite{nasiri2024experimental}.

A phenomenon resembling counter-jet formation can also occur during the impact of a droplet on highly non-wettable surfaces. In this scenario, part of the droplet's kinetic energy is temporarily stored as surface energy upon impact. As the droplet begins to retract, this stored energy drives an upward motion at the droplet's center. The initial impact also excites capillary waves that travel along the droplet interface and converge toward the axis of symmetry, producing a pronounced flow-focusing effect. This mechanism ultimately generates a narrow, upward jet commonly referred to as a Worthington jet \cite{zeff2000singularity}. Such jets are typically elongated and prone to breakup due to the Rayleigh--Plateau instability, often leading to the ejection of one or more satellite droplets \cite{gordillo2020impulsive,michon2017jet,brasz2018minimum}. For liquids with low viscosity, such as water, the Worthington jet can be highly energetic. In extreme cases, the strong upward motion and associated momentum transfer between the droplet and the surface can produce a reaction force sufficient to cause the droplet to completely rebound from the substrate \cite{zhang2022impact}.

\REV{\CUT{To the best of our knowledge, there is no prior study on the impact of hollow droplets with polymer on the surface.}} \MREV{Counter-jet formation, maximum spreading, and deposition regimes are investigated in this research.} We explore experimentally the impact of a hollow droplet with different polymer concentrations on an aluminum surface. This allows us to capture the dynamics of a viscoelastic hollow droplet impact on the surface.

The remainder of this paper is organized as follows: Section~\ref{sec:setup} presents the experimental setup and \NREV{rheological properties of the polymer solutions} at different concentrations. Section~\ref{sec:results} discusses flow dynamics of dense and hollow droplets at different concentrations. Finally, Section~\ref{sec:conclusion} summarizes our findings.

\section{\NREV{Experimental setup and measurements}}
\label{sec:setup}

\begin{figure}[htbp]
	\centering
	\includegraphics[width=0.95\textwidth]{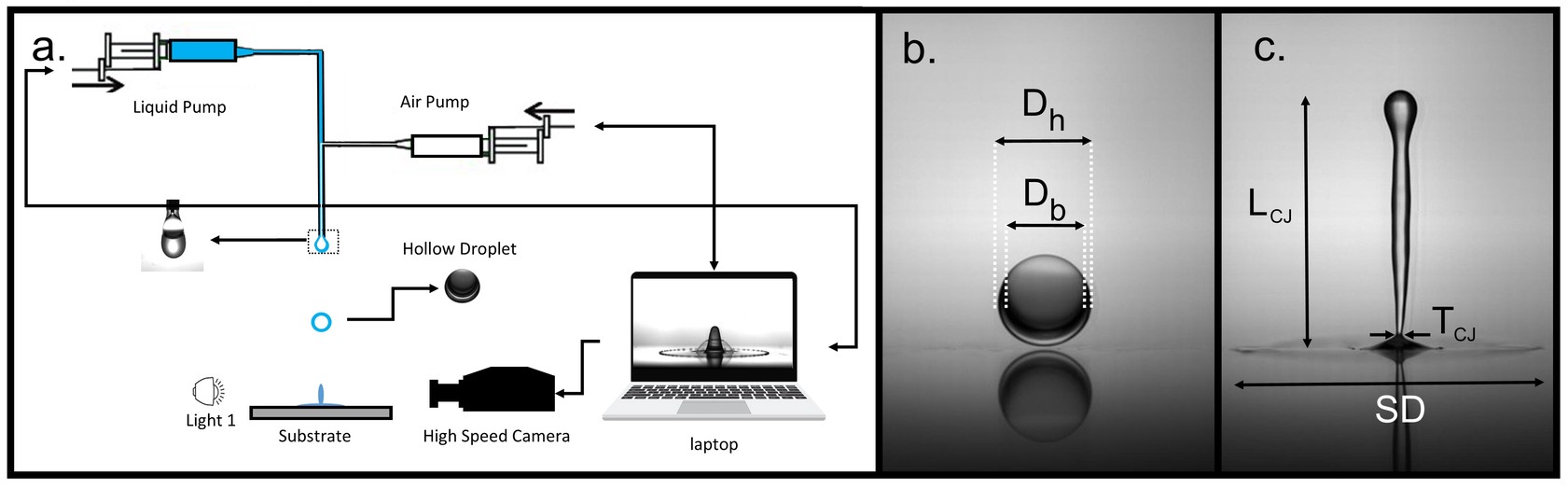}
	\caption{(a) The schematic of the experimental setup of hollow droplet impact.
	\REV{(b) Hollow droplet before impact with hollow droplet diameter ($\Dh$) and bubble diameter ($\Db$). (c) Hollow droplet after impact with the length of counter-jet ($L_{CJ}$) and thickness of the counter-jet $T_{CJ}$.}}
	\label{fig:setup}
\end{figure}

The experimental setup is depicted schematically in Figure~\ref{fig:setup}. \NREV{The impact trials were carried out under ambient laboratory conditions, at room temperature and a relative humidity of 40\%.} \REV{The droplets are injected at a flow rate of $50\ \mu\mathrm{L/min}$ through a needle with an outer diameter (OD) of $2.1$~mm and inner diameter (ID) of $1.7$~mm coupled to a syringe pump (Pico Plus, Harvard Apparatus). To inject air into the liquid droplet, another needle was placed within the primary needle with OD of $0.45$~mm and ID of $0.25$~mm.} \MREV{Air was introduced through the inner needle into the pendant liquid drop before the hollow drop detached from the outer needle.} The needle's height above the surface was changed from \FIXCUT{20 to 70} \FIX{100 to 730}~mm, resulting in impact velocities ranging from \FIXCUT{1.0 to 3.6} \FIX{1.4 to 3.8}~m/s. A high-speed camera (\REVCUT{Phantom v711, Vision Research} \REV{Photron SA1 Fastcam}) \NREV{recorded the droplet impact at 5000~fps with a resolution of $1024\times1024$~pixels and an exposure time of $30\ \mu$s}. One light source was utilized to illuminate the impact spot in order to capture the side view of the droplet impact. The video signal was saved to a PC's memory, and the images were processed with ImageJ software (version 1.46, National Institutes of Health, Bethesda, MD).

{Equation~(\ref{eq:Dbubble}) could be used to \MREV{estimate the bubble diameter and the liquid mass}, considering the fisheye effect.
\begin{equation}
\label{eq:Dbubble}
\Db = 0.86\, D_{b-image}\, .
\end{equation}}
$D_{b-image}$ indicates the measured diameter of the bubble, whereas $\Db$ represents the \MREV{corrected estimate of the bubble diameter}. \MREV{The liquid mass of the hollow droplet was estimated using the measured outer diameter} of the hollow droplet ($\Dh$) and the diameter of the entrapped bubble ($\Db$). \REV{The equivalent diameter of the hollow droplet ($\Deq$) has been calculated based on $\Dh$ and $\Db$ and has been used \NEW{as the characteristic length scale [Eq.~(\ref{eq:Deq})].} The detailed information can be found in the authors' previous work \cite{nasiri2021hollow}. It should be noted that the hollow droplet production is not a totally controlled process, and the sizes of droplets and bubbles differ each time, however, to have a better comparison, only droplets with less than 20\% tolerance in $\Db$ or $\Dh$ are considered. \MREV{Ten droplets were used to characterize the nominal size. For $\Dh=5.6$~mm and $\Db=4.5$~mm, $\Db/\Dh\approx0.804$ and $\Deq\approx4.39$~mm. This normalization does not guarantee equal liquid masses in every trial. The bubble-to-drop diameter ratio was not independently controlled, and its condition-specific variation was not quantified.} Furthermore, in order to facilitate the comparison of the experimental figures, only images of the droplet for almost the same $\Db$ and $\Dh$ sizes are used to produce the figures.}
\NEW{\begin{equation}
\label{eq:Deq}
\REV{\Deq^{3} = \Dh^{3} - \Db^{3}}
\end{equation}}

\NEW{\MREV{The representative impact sequences and spreading model use this equivalent diameter; the diameter convention for the broader regime survey is specified in Section~\ref{sec:regime}.} The key control parameter of the spreading phase is the Weber number, the ratio of the inertial to the capillary forces,
\begin{equation}
\label{eq:We}
\We = \frac{\rho\, U_{0}^{2}\, \Deq}{\gamma}\, ,
\end{equation}}
where $\rho$ is the density of the fluid, $U_{0}$ the impact velocity and $\gamma$ the surface tension coefficient. Another important dimensionless parameter is the Ohnesorge number, which quantifies the ratio of the inertia-capillary to inertia-viscous timescales,
\NEW{\begin{equation}
\label{eq:Oh}
\Oh = \frac{\eta_{s}}{\sqrt{\rho\, \Deq\, \gamma}}\, ,
\end{equation}
where $\eta_{s}$ is the solvent viscosity. Further, in the context of viscoelastic liquids, one may define the Deborah number as
\begin{equation}
\label{eq:De}
\De = \frac{\lambda}{\tau}\, ,
\end{equation}
to compare the elastic response with the process time scale. Here $\lambda$ is the relaxation time of the polymer and $\tau = \Deq/U_{0}$ is the process timescale.}

\MREV{Aqueous poly(ethylene oxide) (PEO) solutions with a nominal molecular mass of $600~\mathrm{kg\,mol^{-1}}$ were investigated at mass concentrations $c_m=0.05$, 0.10, 0.25, 0.50, and 0.75~wt.\%. Their physical and rheological properties are summarized in Table~\ref{tab:fluids}. The reference extensional relaxation time, $\lambda$, was determined using dripping-onto-substrate (DoS) rheometry~\cite{dinic2017pinch}. In this method, a dispensed drop contacts a substrate, forming a liquid bridge whose capillary-driven thinning is recorded. The relaxation time is inferred from the approximately exponential thinning interval.} \REV{A 6061 aluminum surface was used as the impact substrate.} \MREV{A native length-per-pixel calibration and quantitative surface roughness are not established in the present analysis. The acquisition dimensions alone do not determine the resolution of the thinnest filaments; these limitations also restrict comparisons with other substrates.}

\MREV{The adopted image-based measurement bounds are $\pm0.3$~mm for droplet diameter and $\pm0.1$~m/s for impact velocity. These account approximately for spatial calibration, interface identification, and finite image resolution. The parameter bounds shown in the regime map are propagated from these estimates as described in Section~\ref{sec:regime}; they do not include uncertainty in the assumed fluid properties.}

\MREV{The regime survey contains 70 individual impacts at 18 combinations of concentration and release height, with two to eight impacts per combination. The image sequences and time histories presented here are representative measurements. The survey points are individual observations, and their error bars are parameter bounds rather than standard deviations of repeated impacts.}

\NEW{Table~\ref{tab:fluids} summarizes the fluid characteristics used in these experiments. \MREV{The water reference values $\rho=998$~kg/m$^{3}$, $\eta_{s}=0.00089$~Pa\,s, and $\gamma=0.072$~N/m are used in the analysis. } The polymeric contribution to the zero-shear viscosity is $\eta_{p}=\eta_{0}-\eta_{s}$.}

\begin{table}[htbp]
	\caption{Physical properties of the employed fluids. \MREV{DoS relaxation times are means $\pm$ sample standard deviations of five fitted records per polymer concentration. These standard deviations describe the fit-to-fit spread, not the uncertainty from independent solution preparations.}}
	\label{tab:fluids}
	\begin{ruledtabular}
	\begin{tabular}{lccc}
	Liquid & $\eta_{p}/\eta_{s}$ & \rev{$\eta_{p}+\eta_{s}$ (Pa\,s)} & \MREV{$\lambda$ (ms)}\\
	\colrule
	Water (\ed{$c_m=$} 0\,\%)   & 0    & \gr{0.00089} & 0 \\
	Poly.\ (\ed{$c_m=$} 0.05\%) & 0.33 & \gr{0.00118} & \MREV{$0.391\pm0.015$} \\
	Poly.\ (\ed{$c_m=$} 0.10\%) & 0.52 & \gr{0.00135} & \MREV{$0.494\pm0.054$} \\
	Poly.\ (\ed{$c_m=$} \LANGCUT{0.30} \LANG{0.25}\%) & \LANGCUT{1.99} \LANG{1.75} & \gr{\LANGCUT{0.00266} \LANG{0.00245}} & \MREV{$0.861\pm0.023$} \\
	Poly.\ (\ed{$c_m=$} 0.50\%) & 4.40 & \gr{0.00481} & \MREV{$1.394\pm0.021$} \\
	Poly.\ (\ed{$c_m=$} 0.75\%) & 9.66 & \gr{0.00949} & \MREV{$2.049\pm0.011$} \\
	\end{tabular}
	\end{ruledtabular}
\end{table}

\section{Results and discussion}
\label{sec:results}

\subsection{\NREV{Newtonian Droplet Impact}}
\label{sec:newtonian}

\begin{figure}[htbp]
	\centering
	\includegraphics[width=0.95\textwidth]{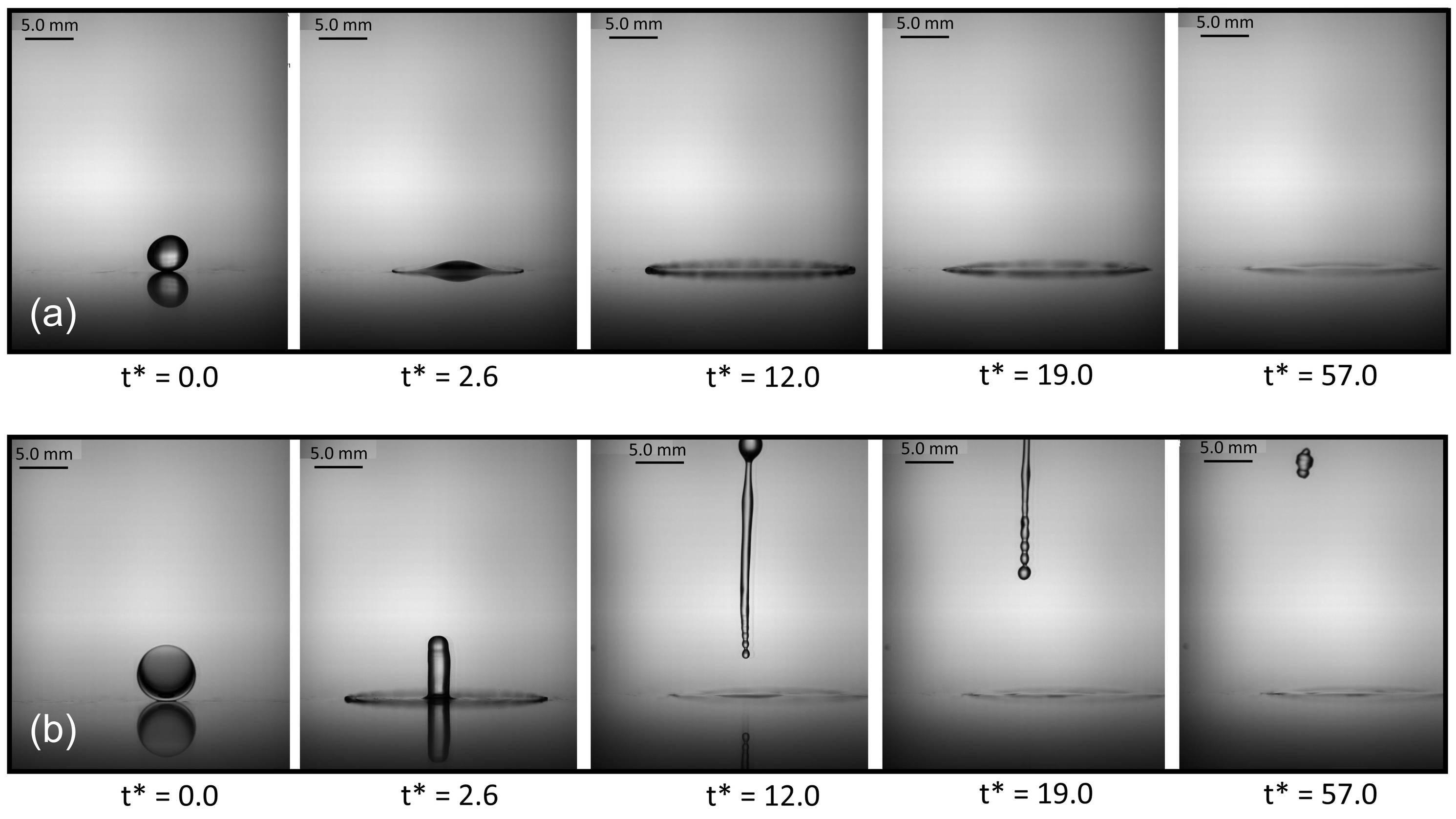}
	\caption{Selected snapshots showing distilled water droplet impact on an aluminum surface at $U_{0}=3.8$~m/s, $\We\approx880$, $\Oh\approx0.0016$, $\De=0$ \NREV{at different dimensionless times} ($\ts = t\,U_{0}/\Deq$).
	(a) A dense water droplet with $\Deq=4.4$~mm. (b) A hollow droplet impact with $\Dh=5.6$~mm, $\Db=4.5$~mm and $\Deq=4.4$~mm.}
	\label{fig:water}
\end{figure}

Figure~\ref{fig:water} shows selected snapshots of distilled water droplet impacting at $U_{0}=3.8$~m/s on an aluminum surface at different time steps ($\ts = t\,U_{0}/\Deq$). The \REV{flattening} of the dense and hollow droplets \NREV{is shown in} \NEW{Figures~\ref{fig:water}(a)} and \ref{fig:water}(b), respectively.

\NREV{When a dense droplet impacts a solid surface, its inertia produces a rapid rise in pressure near contact and drives radial spreading [Figure~\ref{fig:water}(a) at $\ts=0$]. The drop spreads smoothly with limited surface disturbances at $\ts=2.6$ and reaches its maximum spreading diameter at $\ts\approx4$ [Figure~\ref{fig:SDdense}]. The liquid sheet then retracts under surface tension. Small finger-like rim disturbances visible at $\ts=12$ merge into a more continuous sheet by $\ts=19$; the $\ts=57$ snapshot shows a later stage. The droplet gradually approaches an equilibrium configuration determined by wetting and the balance of capillary and gravitational forces.}

In contrast, the impact of a hollow droplet on a solid surface leads to the formation of a counter-jet that emerges perpendicular to the surface \NREV{as soon as the droplet impacts and begins to spread and flatten} [Figure~\ref{fig:water}(b)]. This counter-jet develops due to a pressure differential inside the droplet, which is created by an entrapped air bubble during impact. As the hollow droplet continues to spread, the central counter-jet keeps rising [Figure~\ref{fig:water}(b) at $\ts=2.6$] and eventually detaches from the liquid sheet, carrying away part of the liquid mass from the surface. \FIXCUT{The Kelvin--Helmholtz instability along the interface of water and air with different tangential velocities and densities causes the formation of vortices and waves along the detached counter-jet which breaks up into several droplets} \FIX{The detached counter-jet breaks up into several droplets through the Rayleigh--Plateau instability} [Figure~\ref{fig:water}(b) at $\ts=19$--$57$].

\CUT{The entrapped bubble plays a significant role in the droplet's deformation behavior. During the early stages of impact, the bubble undergoes compression and subsequent collapse under the high pressure exerted by the droplet. The bursting of the bubble produces waves and disturbances on the liquid sheet, disrupting its stability. This disturbance often causes the sheet to rupture at its center, leading to the formation of a hole within the spreading film. Consequently, when the liquid begins to recoil toward the center, it fails to reform into a hemispherical shape. Instead, it stabilizes into a donut-like configuration on the surface. The generation of a counter-jet, bubble breakup, and the associated surface perturbations are characteristic behaviors of hollow droplet impacts and are consistently observed across surfaces with varying wettability \cite{nasiri2024experimental}.}

\CUT{A phenomenon resembling counter-jet formation can also occur during the impact of a droplet on highly non-wettable surfaces. In this scenario, part of the droplet's kinetic energy is temporarily stored as surface energy upon impact. As the droplet begins to retract, this stored energy drives an upward motion at the droplet's center. The initial impact also excites capillary waves that travel along the droplet interface and converge toward the axis of symmetry, producing a pronounced flow-focusing effect. This mechanism ultimately generates a narrow, upward jet commonly referred to as a Worthington jet \cite{zeff2000singularity}. Such jets are typically elongated and prone to breakup due to the Rayleigh--Plateau instability, often leading to the ejection of one or more satellite droplets \cite{gordillo2020impulsive,michon2017jet,brasz2018minimum}. For liquids with low viscosity, such as water, the Worthington jet can be highly energetic. In extreme cases, the strong upward motion and associated momentum transfer between the droplet and the surface can produce a reaction force sufficient to cause the droplet to completely rebound from the substrate \cite{zhang2022impact}.}

\subsection{Polymeric Droplets}
\label{sec:polymeric}

\begin{figure}[htbp]
	\centering
	\includegraphics[width=0.92\textwidth]{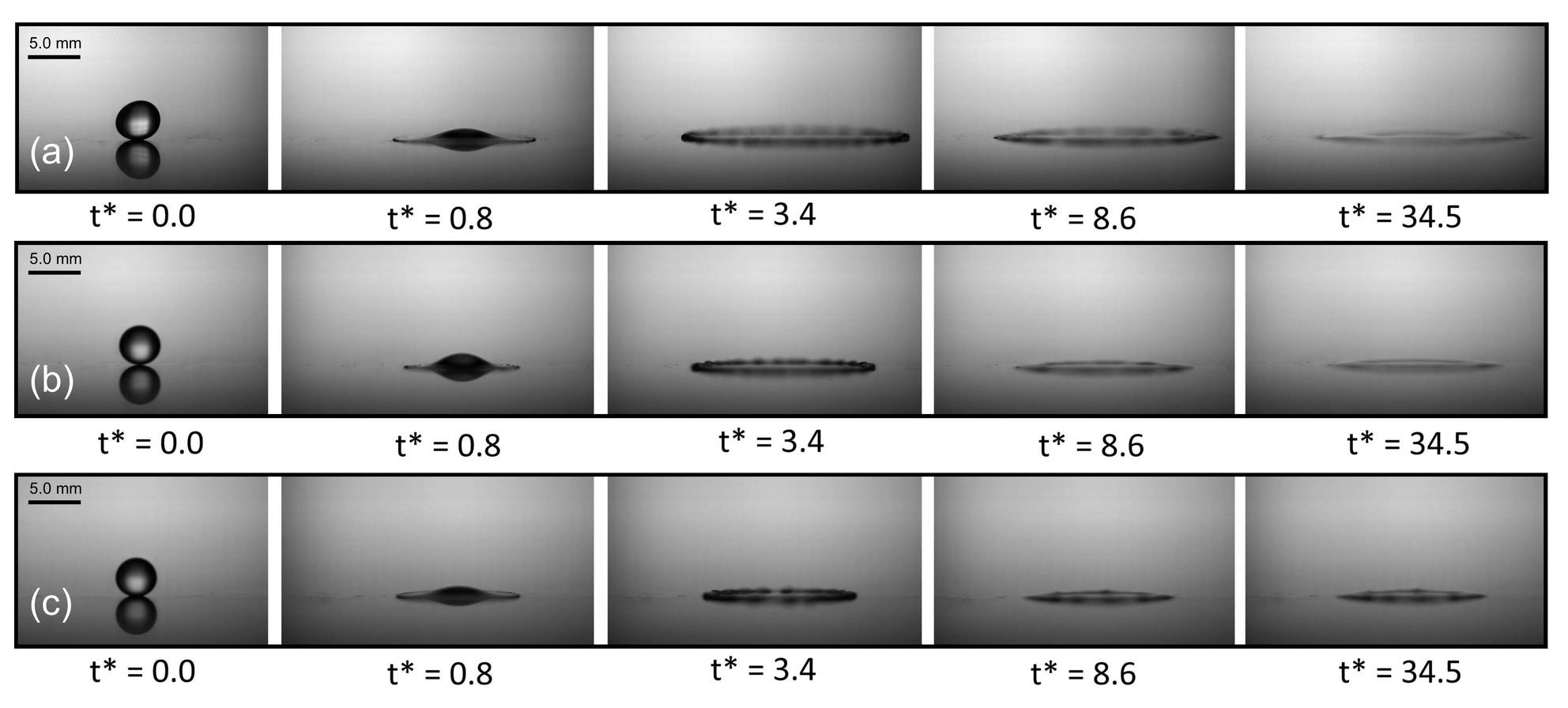}
	\caption{Selected snapshots showing a dense droplet impact on an aluminum surface at $U_{0}=3.8$~m/s, $\We\approx880$, $\Oh\approx0.0016$.
	(a) A water droplet with 0.05\% polymer. (b) A water droplet with \LANGCUT{0.3} \LANG{0.25}\% polymer. (c) A water \NEW{droplet} with 0.75\% polymer.}
	\label{fig:densePolymer}
\end{figure}

Figure~\ref{fig:densePolymer} illustrates a series of droplets with different polymeric concentrations, $0.05\% \le c_{m} \le 0.75\%$. The impact velocity along this set of experiments is the same for these cases. The droplet \REV{flattening} along different polymeric concentrations shows a qualitatively similar development. However, as $c_{m}$ increases, \ed{the maximum spreading diameter decreases.} \CUT{We
		hypothesize this behavior is related to the increase of energy dissipation due to elevation of
		apparent viscosity of the fluid. }

\begin{figure}[htbp]
	\centering
	\includegraphics[width=0.7\textwidth]{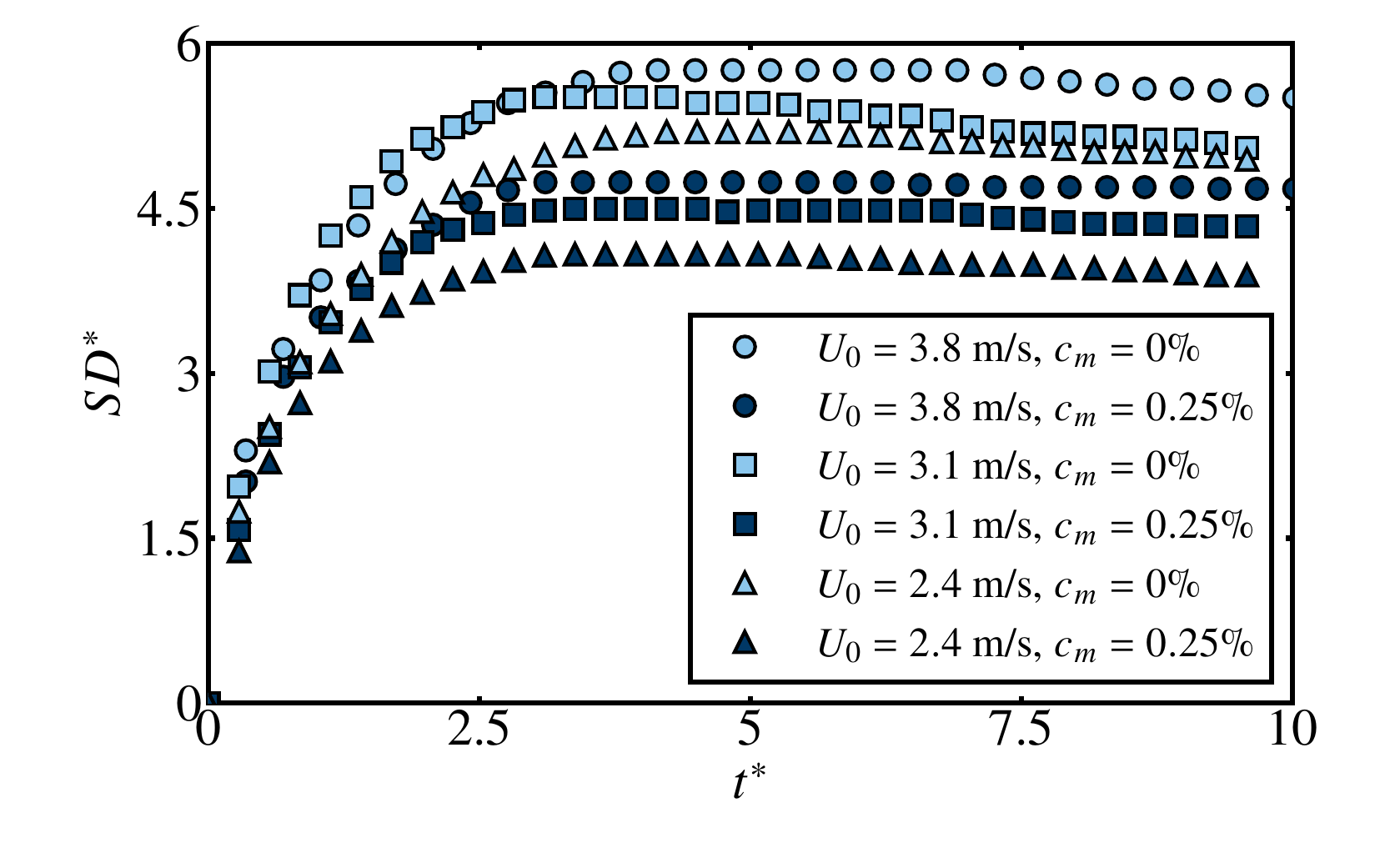}
	\caption{Dimensionless spreading diameter ($SD^{*} = SD/\Deq$) versus dimensionless time ($\ts = t\,U_{0}/\Deq$) \MREV{for dense water and 0.25\% PEO droplets at different impact velocities.}}
	\label{fig:SDdense}
\end{figure}

Figure~\ref{fig:SDdense} \NEW{quantitatively presents} the development of spreading diameter ($SD^{*} = SD/\Deq$) at different $c_{m}$ and $U_0$. As \NEW{expected, higher impact velocity will lead to larger} spreading diameter. On the \NEW{other hand, increase} of $c_m$ \NEW{in each impact velocity} decreases \NEW{the spreading diameter. \CUT{Furthermore, the required time of reaching the maximum spreading diameter also decreases with $c_{m}$ increment.}
\NREV{This reduction is consistent with energy taken up by polymer deformation and relaxation.} The spreading lamella stretches the polymer chains, and the work done on them is withdrawn from the kinetic energy that drives the spreading: part of it is stored elastically in the stretched chains and part is dissipated as they relax. This contribution grows} with $c_{m}$, \NEW{so progressively less energy is left to extend the lamella.}

\begin{figure}[htbp]
	\centering
	\includegraphics[width=0.95\textwidth]{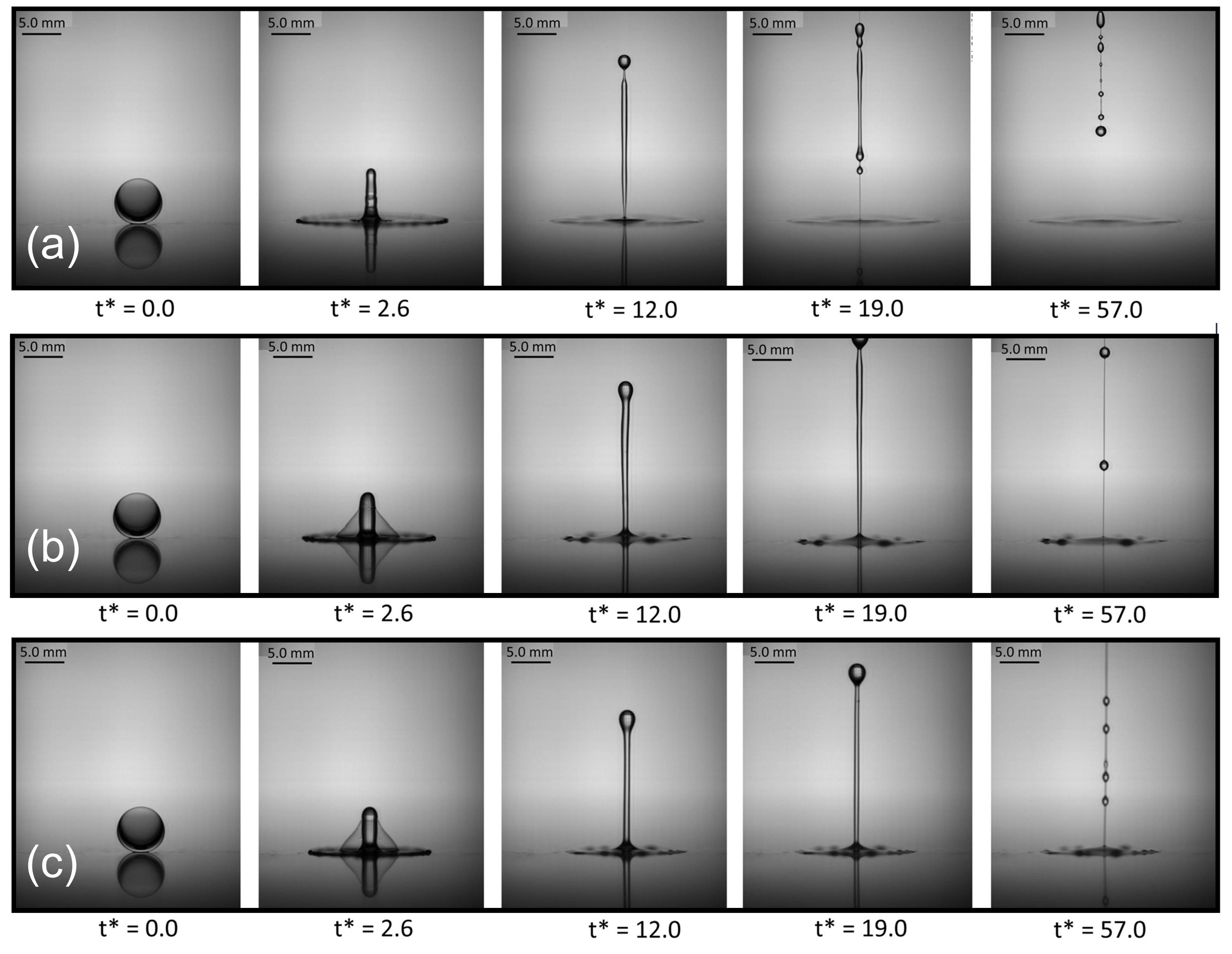}
	\caption{Selected snapshots showing a hollow droplet impact on an aluminum surface at $U_{0}=3.8$~m/s, $\We\approx880$, $\Oh\approx0.0016$.
	(a) A hollow water droplet with 0.05\% polymer. (b) A water droplet with \LANGCUT{0.3} \LANG{0.25}\% polymer. (c) A water \NEW{droplet} with 0.75\% polymer.}
	\label{fig:hollowPolymer}
\end{figure}

\begin{figure}[htbp]
	\centering
	\subfloat[\label{fig:spreadChar_a}]{
		\includegraphics[trim=1cm 1cm 0.5cm 0.4cm, clip=true,width=0.48\textwidth]{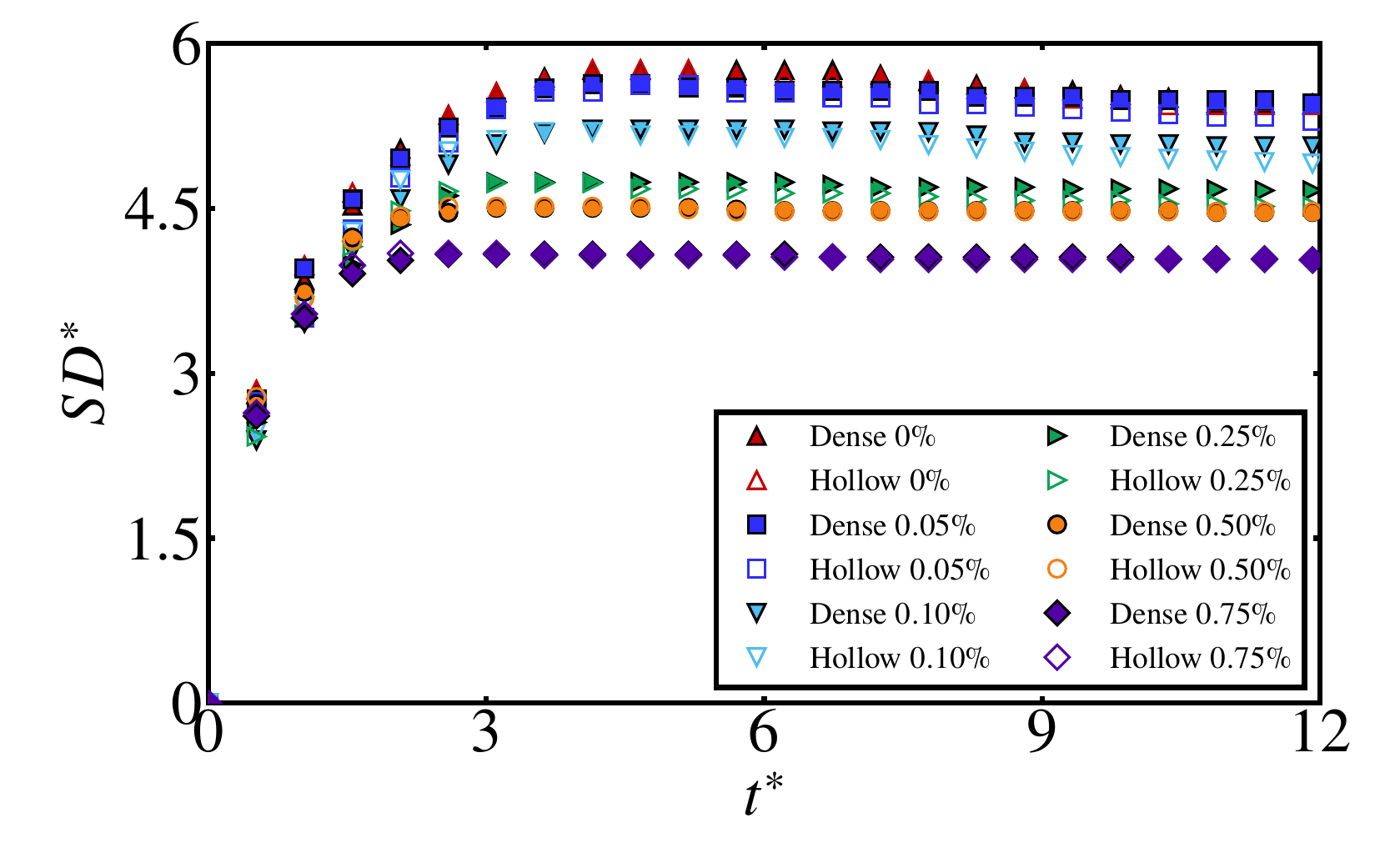}
	}\hfill
	\subfloat[\label{fig:spreadChar_b}]{
		\includegraphics[trim=1cm 1cm 0.5cm 0.4cm, clip=true,width=0.48\textwidth]{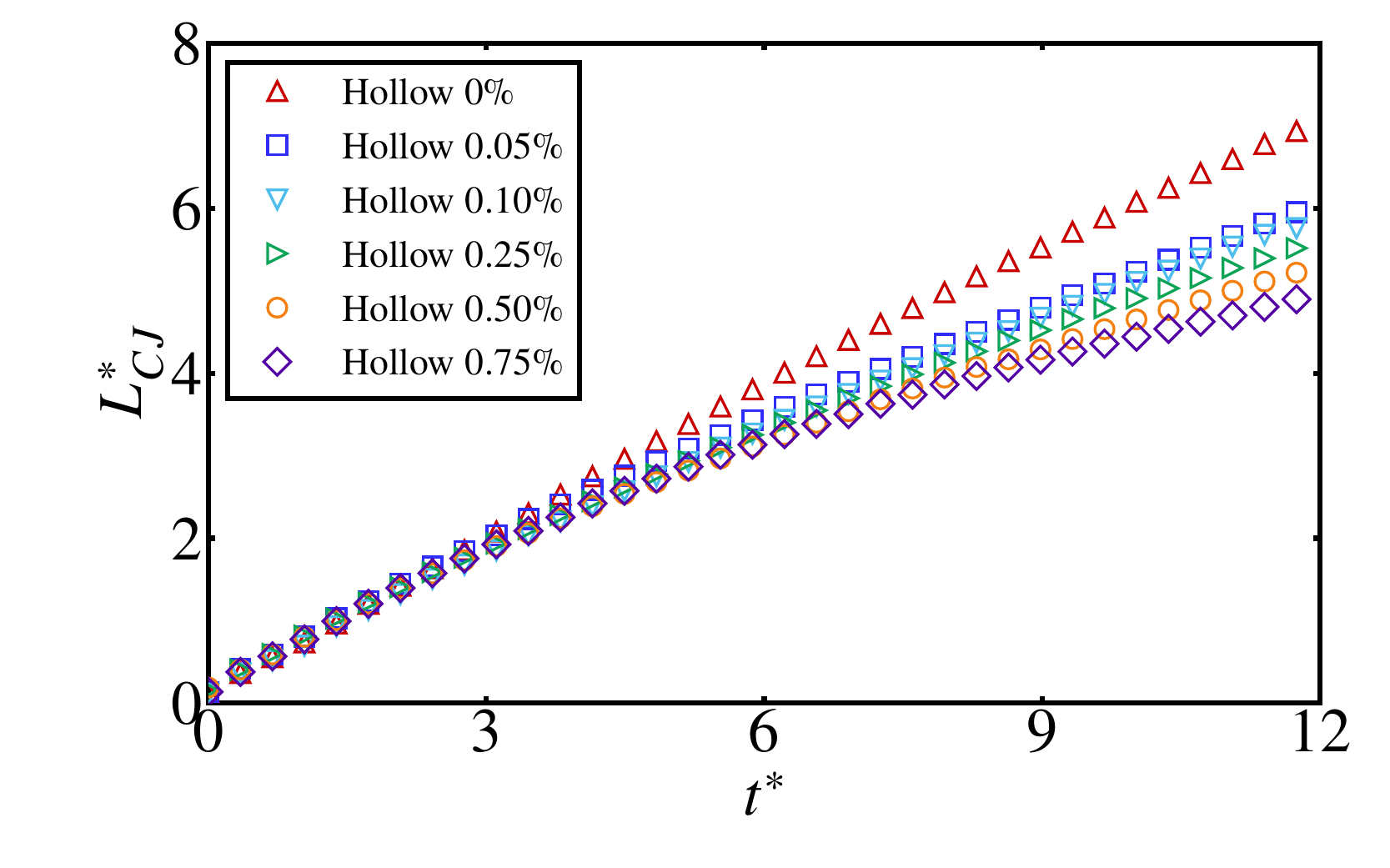}
	}
	\caption{Spreading characteristics of \NREV{droplets at different polymer concentrations, with} $\Deq=4.4$~mm and $U_{0}=3.8$~m/s ($\We\approx880$) impacting on a substrate.
	(a) Dimensionless spreading diameter ($SD^{*} = SD/\Deq$) versus dimensionless time ($\ts = t\,U_{0}/\Deq$) for dense and hollow droplets ($\Dh=5.6$~mm, $\Db=4.5$~mm and $\Deq=4.4$~mm). (b) Dimensionless counter-jet \NEW{length ($L^{*}_{CJ} = L_{CJ}/\Deq$)} versus dimensionless time for the hollow droplet.
	$\Dh$, $\Db$ and $\Deq$ are the initial diameter of the hollow droplet, the initial diameter of bubble inside hollow droplet and equivalent diameter of a dense droplet with the same mass of a hollow droplet, respectively.}
	\label{fig:spreadChar}
\end{figure}

\NREV{Figure~\ref{fig:hollowPolymer} shows snapshots of hollow droplet impacts at $c_m=0.05$, 0.25, and 0.75\%. Maximum spreading occurs at early times. As the polymer concentration increases, the spreading diameter decreases (see Figure~\ref{fig:hollowPolymer} at $\ts=2.6$), consistent with the trend for dense droplets in Figure~\ref{fig:densePolymer}. Figure~\ref{fig:spreadChar}(a) shows the evolution of the dimensionless spreading diameter for dense and hollow droplets at different polymer concentrations.}
\CUT{Figure~\ref{fig:spreadChar}(a) also confirms that, in hollow droplets, the increase of polymeric concentration monotonically decreases the spreading diameter.}
During the initial kinematic phase, spreading is similar for all fluids ($\ts \approx 2$) \cite{balzan2021drop}. At this stage there are minor differences in the droplet shape which persist throughout spreading. After $\ts \approx 2$, the droplet spreading differences become more noticeable. At $\ts \approx 3$, the water droplet continues to spread, \MREV{while the polymeric traces approach maximum spreading in Figure~\ref{fig:spreadChar_a}.} \NREV{While droplets of the other solutions} (except, marginally, 0.05\% solution) start to retract, the water droplets continue to thin and spread \cite{balzan2021drop}. Although variations in polymer concentration do not induce qualitative differences in the overall morphology during the spreading stage, they do affect the rate of spreading and counter-jet development. As the droplet continues to flatten, there is a reduction in spreading rate with increasing viscosity as more energy is dissipated. \CUT{We further compare the spreading diameter of dense and hollow droplets at various polymeric concentrations in Figure~\ref{fig:spreadChar}(a).} \NEW{Furthermore,} at a given concentration, the evolution of spreading diameter remains similar in both dense and hollow droplets. In our previous works \cite{nasiri2024experimental,nasiri2021hollow,nasiri2023flattening}, we attributed this behavior to the stored energy, bubble rupture, and the induced perturbations by the \NEW{\REV{ruptures}.}

The other difference is the presence of the bubble shell and the counter-jet at the same time on liquid with higher concentrations. We assume the slower advancement of the counter-jet in higher concentrations and elasticity of the fluid are associated with this behavior. Nevertheless, as the counter-jet continues to grow, the bubble shell cannot endure the tension and eventually breaks up. As mentioned, \NREV{the counter-jet height is another parameter that differs}. As is demonstrated, \NREV{the counter-jet at the lowest concentration rises higher than those at higher concentrations} (see Figure~\ref{fig:hollowPolymer} at $\ts=12$).
Figure \ref{fig:spreadChar_b} \NEW{presents the evolution of dimensionless counter-jet length.
 Figure~\ref{fig:spreadChar}(b)} also shows the monotonic decrease of counter-jet height with increase of the polymer concentration.
 \MREV{For comparable nominal liquid volumes and impact velocities, the decrease of counter-jet height could be associated with} \CUT{higher viscous dissipation in higher concentrations} \NEW{the polymer contribution to the energy of the impact. The rising jet stretches the polymer chains, and the work done on them is withdrawn from the kinetic energy that drives the jet upward: part of it is stored elastically} in \NEW{the stretched chains and part is dissipated as they relax. \NREV{A larger polymer contribution at higher concentration could therefore limit the jet rise.}}

In contrast to the counter-jet height, the thickness of the counter-jet increases with polymer concentration. \FIXCUT{The Newtonian droplet} \FIX{The lowest-concentration (0.05\%) droplet} \MREV{remains connected at $\ts\approx12$ and 19 and is detached in the later $\ts\approx57$ frame [Figure~\ref{fig:hollowPolymer}(a)], while the higher-concentration filaments remain attached [Figures~\ref{fig:hollowPolymer}(b) and \ref{fig:hollowPolymer}(c)].} As time progresses and the counter-jet rises, the thickness of the attached droplet gradually decreases (see \NEW{Figures~\ref{fig:hollowPolymer}(b)} and \ref{fig:hollowPolymer}(c) \CUT{and \ref{fig:hollowPolymer}(d)} at $\ts \approx 19$). This filament thinning results from the interplay between capillary forces, which promote necking, and elastic stresses, which resist thinning by stretching the polymers along the flow direction. The resulting elastic tension counteracts capillary pressure, leading to the formation of a long, \MREV{persistent} cylindrical thread rather than rapid pinch-off as seen in Newtonian fluids. Eventually, the filament thins \CUT{exponentially} until the droplet with lower polymer concentration can no longer support the tension, causing it to detach [see Figure~\ref{fig:hollowPolymer}(a) at $\ts \approx 57$]. In contrast, droplets with higher polymer concentrations remain attached even at later times [see \NEW{Figures~\ref{fig:hollowPolymer}(b)} and \ref{fig:hollowPolymer}(c) \CUT{and \ref{fig:hollowPolymer}(d)} at $\ts \approx 57$]. The last difference between the hollow water droplet and polymer droplets is the formation of beads-on-a-string (BOAS) structures. \NREV{These structures comprise nearly cylindrical filaments connecting approximately spherical beads [see Figure~\ref{fig:Abreakup} in the Appendix].} These BOAS structures are the result of a balance between capillarity and the viscoelasticity of the polymer \NREV{and were observed here in counter-jets formed by polymer solutions}. Generally, elasticity resists the breakup, shown by a long thin filament. In this case, an instability is seen where surface tension \MREV{reduces interfacial area at fixed liquid volume} by creating spherical interfaces, with smaller thin filaments formed in between each bead. \NREV{The higher-concentration filaments persist for longer before breakup.}

The counter-jet height during spreading time is the other parameter that has been illustrated in Figure~\ref{fig:spreadChar}(b). The counter-jet starts to {form} from the impact time for all solutions. The slope of this diagram is almost identical for all solutions at the initial times of the impact ($\ts \approx 2$). After that, \NREV{differences between the counter-jet growth rates} are more visible. Water counter-jet grows faster as the time proceeds, while, as the concentration increases, the growth rate decreases and the counter-jet formed by the 0.75\% solution \NREV{has the shortest counter-jet height among the cases considered}. \NREV{This trend is consistent with stronger resistance to jet extension from viscous and polymeric stresses at higher concentrations.} In addition, it was observed that the bubble rupture is delayed in the hollow droplet with higher concentrations (see Figure~\ref{fig:Arupture} \NEW{in the Appendix)} \NREV{and interaction with the shell may further alter the counter-jet motion.}

\begin{figure}[htbp]
	\centering
	\subfloat[\label{fig:cjChar_a}]{
		\includegraphics[trim=1cm 1cm 0.5cm 0.4cm, clip=true,width=0.48\textwidth]{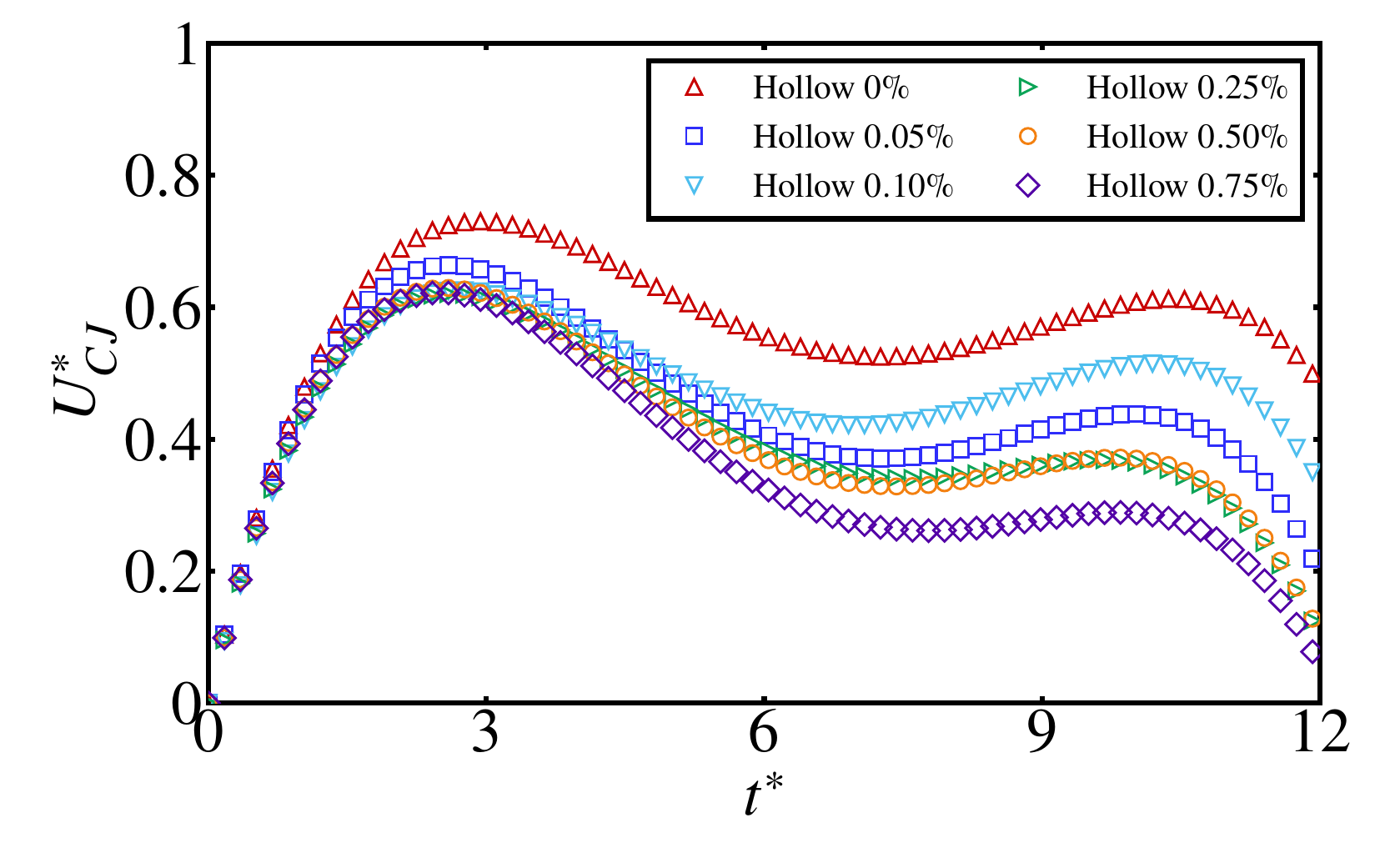}
	}\hfill
	\subfloat[\label{fig:cjChar_b}]{
		\includegraphics[trim=0.2cm 1cm 0.5cm 0.1cm, clip=true,width=0.48\textwidth]{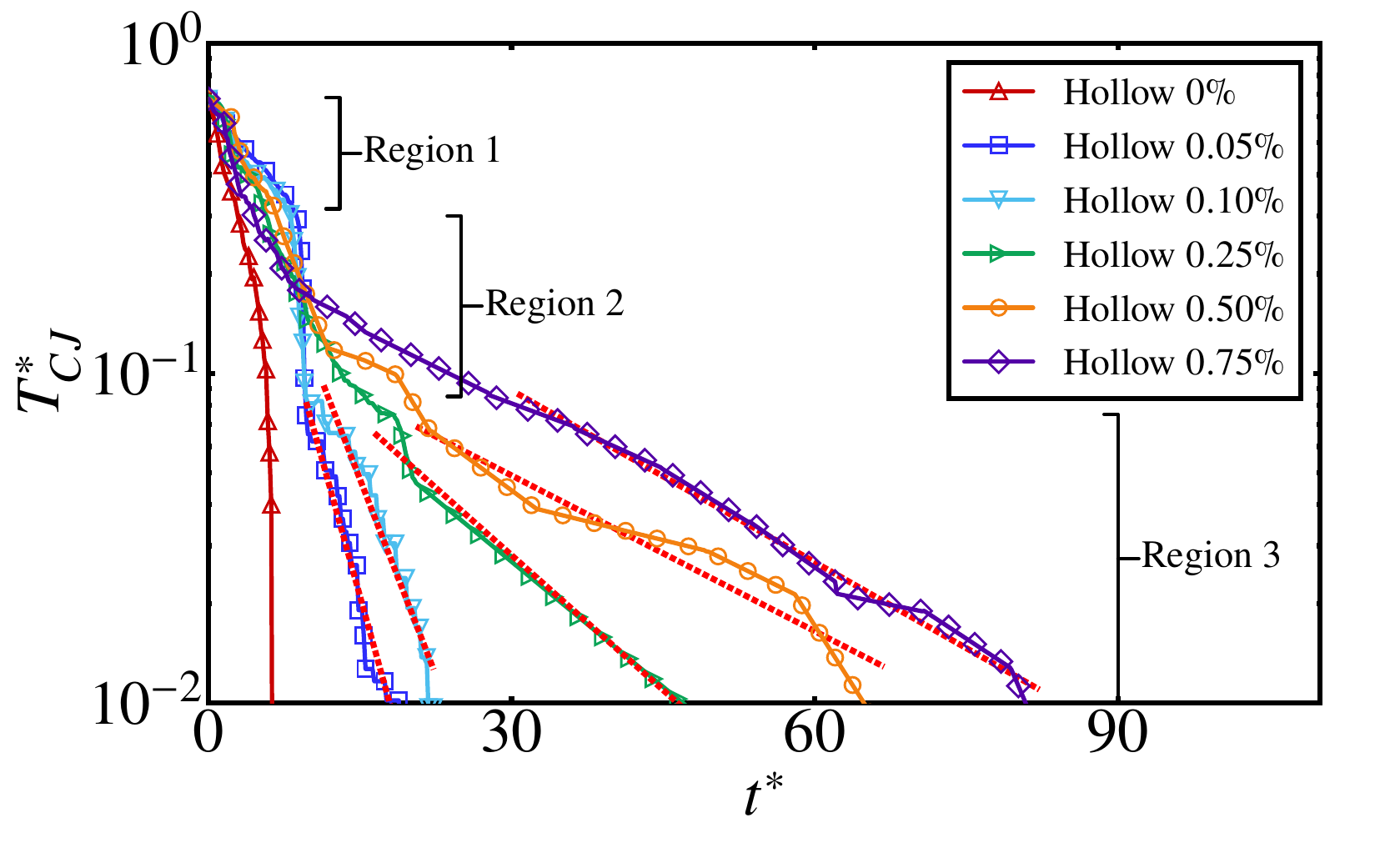}
	}
	\caption{\REV{\MREV{Counter-jet characteristics of polymeric hollow droplets, with water as a reference,} with $\Deq=4.4$~mm and $U_{0}=3.8$~m/s ($\We\approx880$) impacting on a substrate.
	(a) Dimensionless counter-jet velocity ($U_{CJ}^{*} = U_{CJ}/U_{0}$) versus dimensionless time ($\ts = t\,U_{0}/\Deq$) for the hollow droplet.
	(b) Dimensionless counter-jet thickness ($T_{CJ}^{*} = T_{CJ}/\Deq$) versus dimensionless time for the hollow droplet.} \NEW{The red dashed lines in (b) are fits of Eq.~(\ref{eq:thinning}) to Region} 3 \MREV{over $0.008\leq T^{*}_{CJ}\leq0.08$}.}
	\label{fig:cjChar}
\end{figure}

{The trendline of counter-jet tip velocity for different solutions is demonstrated in Figure~\ref{fig:cjChar}(a). As the counter-jet starts to form, the velocity of the tip increases due to the energy of the falling droplet. \MREV{The subsequent peaks and decreases in tip velocity may reflect the interaction with the bubble shell and the evolving jet shape. The tip-velocity histories alone do not quantify the energy transferred during bubble rupture.}}

{To quantitatively compare the extent of filament thinning, Figure~\ref{fig:cjChar}(b) displays the \MREV{counter-jet thickness for the polymeric solutions, with water shown as a reference. The rapid breakup of the Newtonian counter-jet is illustrated in Figure~\ref{fig:water}(b). In polymer solutions, the thickness evolves through the stages described below.}}

{\NREV{The polymeric traces exhibit an initial stage of gradual thinning (Region 1).} This is followed by a sudden reduction in filament thickness (Region 2), which corresponds to the formation of the cylindrical threads during the thinning process. \MREV{Region 3 contains an approximately exponential thinning interval; final pinch-off need not follow the same law.} It is noteworthy that as the polymer concentration increases, Region 2 becomes progressively shorter, and Regions 1 and 2 tend to merge into a single phase of \MREV{gradual thinning} (e.g., for $c_{m}=0.75\%$). It should be mentioned that these regions are defined based on the changes in the rate of counter-jet thickness reduction over time.}

\MREV{The late-stage thinning can be interpreted using an elasto-capillary scaling. For a slender, nearly cylindrical Oldroyd-B filament, a balance between capillary pressure, $\approx2\gamma/T_{CJ}$, and polymeric tensile stress predicts exponential thinning when inertia and solvent-viscous stresses are negligible~\cite{entov1997effect,anna2001elasto,clasen2006beads}:}
\begin{equation}
T^{*}_{CJ}\left(t^{*}\right) \propto \exp\left(-\frac{t^{*}}{3\,\De_{e}}\right),
\qquad
\De_{e}=\frac{\lambda_{e}U_{0}}{\Deq},
\label{eq:thinning}
\end{equation}
\MREV{Here, $\lambda_{e}$ is an apparent relaxation time inferred from the impact filament, distinguished from the reference DoS relaxation time $\lambda$. The red dashed lines in Figure~\ref{fig:cjChar}(b) are unweighted least-squares fits of $\ln T^{*}_{CJ}$ against $t^{*}$ over $0.008\leq T^{*}_{CJ}\leq0.08$, retaining positive measurements at distinct times. For a fitted slope $m$, $\lambda_{e}=-\Deq/(3U_{0}m)$, using $\Deq=4.4$~mm and $U_{0}=3.8$~m/s. Table~\ref{tab:lambdae} reports the resulting times and the coefficients of determination in logarithmic thickness.}

\MREV{The apparent times increase overall from $1.5$~ms at $0.05$\% PEO to approximately $10$~ms at the two highest concentrations, although the trend is not monotonic. Both the impact and DoS estimates arise from extensional thinning, but their different geometries and deformation histories prevent their difference from establishing a unique material relaxation time. Changing the lower fit cutoff from 0.008 to 0.006 or 0.010, with the upper cutoff fixed at 0.08, changes $\lambda_{e}$ by up to approximately 13\%; this is a fit-window sensitivity, not a confidence interval. The exponential fits are consistent with an elasto-capillary interpretation, but do not independently establish the required stress balance or a calibrated impact rheometer. Spatial resolution and filament nonuniformity limit the inference, and finite chain extensibility can terminate the exponential regime before pinch-off~\cite{entov1997effect,dinic2017pinch}.}

\begin{table}[htbp]
	\caption{\MREV{Apparent relaxation times from the fitted interval in Region 3 of Figure~\ref{fig:cjChar}(b), compared with the mean DoS times in Table~\ref{tab:fluids}. Fits use 56, 64, 200, 269, and 300 temporal samples in concentration order; $R^{2}$ is calculated in logarithmic thickness and is not an uncertainty estimate.}}
	\label{tab:lambdae}
	\begin{ruledtabular}
	\begin{tabular}{cccc}
	\tb{$c_{m}$ (\%)} & \tb{$\lambda_{e}$ (ms)} & \tb{$\lambda$ (ms)} & \MREV{$R^{2}$ (log fit)}\\
	\colrule
	\tb{0.05} & \tb{1.5}  & \MREV{0.391} & \tb{0.96} \\
	\tb{0.10} & \tb{2.1}  & \MREV{0.494} & \tb{0.95} \\
	\tb{\LANGCUT{0.30} \LANG{0.25}} & \tb{6.1}  & \MREV{0.861} & \tb{0.99} \\
	\tb{0.50} & \tb{10.7} & \MREV{1.394} & \MREV{0.91} \\
	\tb{0.75} & \MREV{9.6}  & \MREV{2.049} & \tb{0.99} \\
	\end{tabular}
	\end{ruledtabular}
\end{table}

\CUT{The viscosity of shear-thinning liquids increases as the shear rate decreases, whereas in Newtonian fluids, viscosity remains constant regardless of shear rate. In the viscosity curve of a shear-thinning fluid, two Newtonian plateaus can be identified: a maximum viscosity at very low shear (zero-shear viscosity, \mbox{$\eta_{0}$}), and a minimum viscosity at very high shear rates (infinite shear viscosity, \mbox{$\eta_{\infty}$}). Between these two limits, reducing the shear rate leads to an increase in viscosity \mbox{\cite{richard2002contact}}. This behavior can be explained by the interplay between energy dissipation, caused by viscous forces, and energy storage, associated with the elastic response of polymer chains. First, the polymers in the solution resist droplet spreading, and increasing their concentration reduces the maximum spreading distance of the droplet. Second, as the apparent viscosity rises, more kinetic energy is dissipated during spreading, further limiting the droplet's maximum spread.}

\CUT{The other parameter that is affected by polymer concentration is the breakup time of the counter-jet from the surface. Figure~\ref{fig:Abreakup} shows selected snapshots of counter-jet breakup after a hollow droplet impact on an aluminum surface. For a water hollow droplet [Figure~\ref{fig:Abreakup}(a)], the breakup of the counter-jet occurs in only $2\Delta t$ ($\Delta t = 0.0002$~ms), this is while the breakup time increases by polymer concentration increment. This breakup time become $25\Delta t$, $75\Delta t$, and $400\Delta t$ for solution with concentration of 0.05\%, 0.10\% and 0.5\%, respectively.}

\CUT{A distinct behavior visible in Figure~\ref{fig:Abreakup} is the emergence of beads-on-a-string (BOAS) patterns. In this configuration, nearly cylindrical threads link a sequence of droplet-like beads, creating a pearl-necklace-type appearance. Such BOAS patterns arise from the interplay between capillary forces, which favor droplet formation, and the elastic response of the polymer chains, which resists filament breakup. This phenomenon is characteristic of counter-jets in viscoelastic polymer solutions. The elastic nature of the fluid delays the rupture of the filament, allowing a long, slender thread to persist between the beads. However, surface tension drives the system toward minimizing surface energy by producing spherical segments, while fine filaments remain as connectors. With increasing polymer concentration, these filaments tend to extend further, become thinner, and resist breakup for a longer period.}

\begin{figure}[htbp]
	\centering
	\subfloat[\label{fig:velChar_a}]{
		\includegraphics[trim=1cm 1cm 0.5cm 0.4cm, clip=true,width=0.48\textwidth]{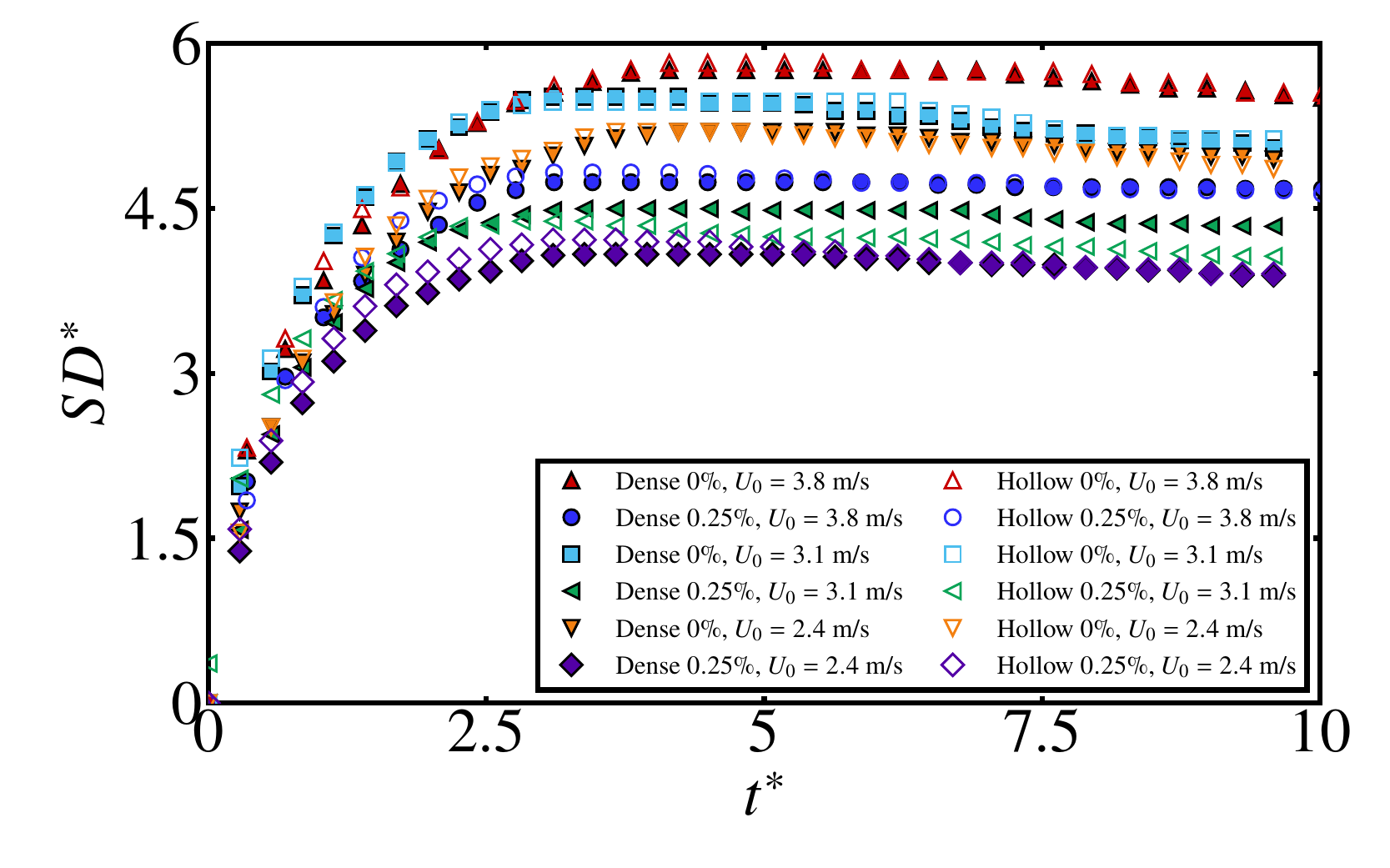}
	}\hfill
	\subfloat[\label{fig:velChar_b}]{
		\includegraphics[trim=1cm 1cm 0.5cm 0.4cm, clip=true,width=0.48\textwidth]{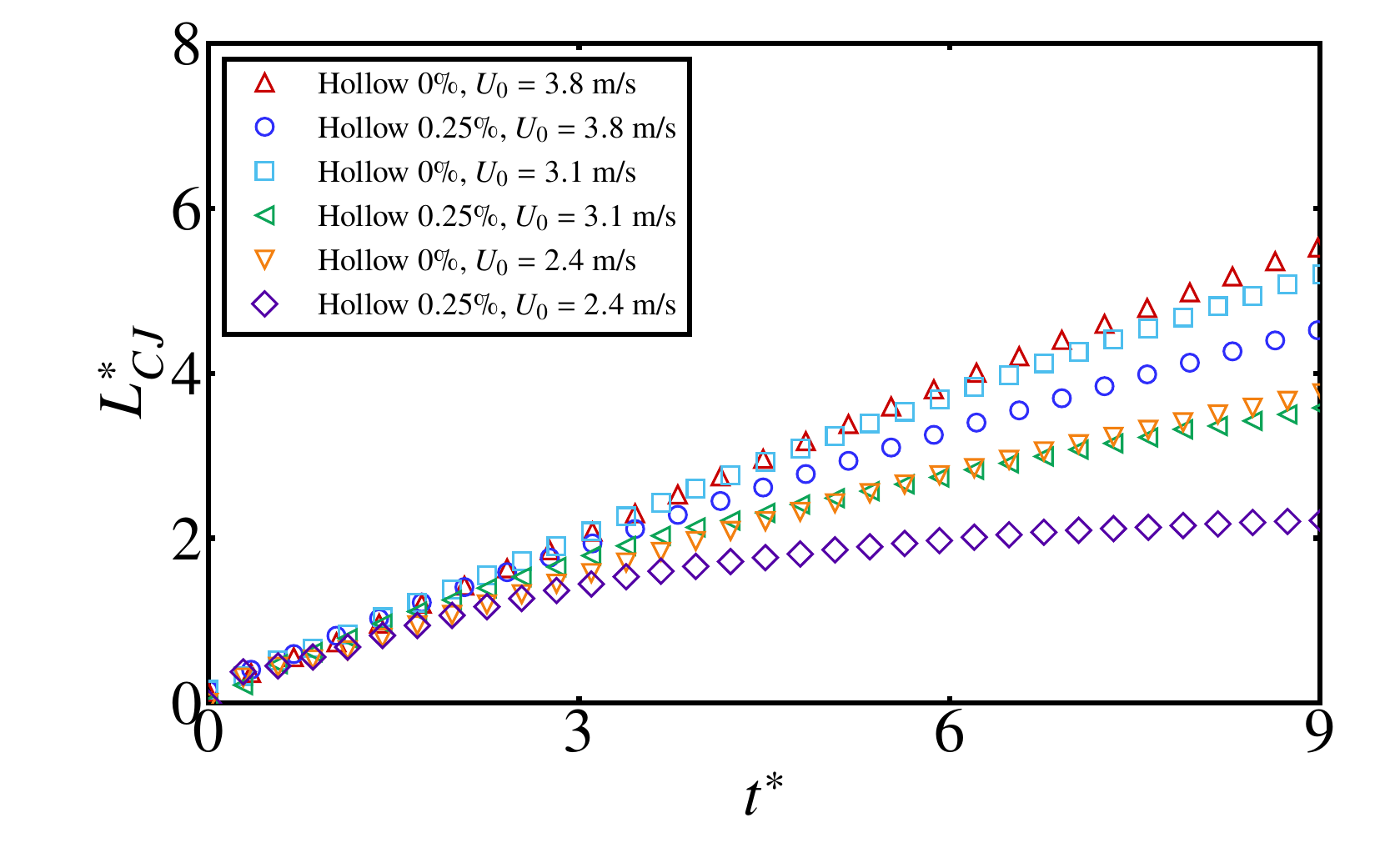}
	}
	\caption{\MREV{Spreading and counter-jet characteristics of water and} \LANGCUT{0.30} \LANG{0.25}\% \MREV{PEO droplets} with $\Deq=4.4$~mm at different impact velocities $U_{0}=2.4$~m/s ($\We\approx350$), $U_{0}=3.1$~m/s ($\We\approx585$), $U_{0}=3.8$~m/s ($\We\approx880$).
	(a) Dimensionless spreading diameter ($SD^{*} = SD/\Deq$) versus dimensionless time ($\ts = t\,U_{0}/\Deq$) for dense and hollow droplets ($\Dh=5.6$~mm, $\Db=4.5$~mm and $\Deq=4.4$~mm). (b) Dimensionless counter-jet \NEW{length ($L^{*}_{CJ} = L_{CJ}/\Deq$)} versus dimensionless time for the hollow droplet.
	$\Dh$, $\Db$ and $\Deq$ are the initial diameter of the hollow droplet, the initial diameter of bubble inside hollow droplet and equivalent diameter of a dense droplet with the same mass of a hollow droplet, respectively.}
	\label{fig:velChar}
\end{figure}

The effect of droplet impact velocity on the flattening of hollow and dense droplets \REV{is} the other parameter that has been investigated (Figure~\ref{fig:velChar}). \MREV{Figure~\ref{fig:velChar} compares water and 0.25\% PEO droplets at the same impact velocities.} \NREV{For a Newtonian droplet, the maximum spreading diameter is governed by the initial kinetic and surface energies, the change in interfacial energy, and viscous dissipation during impact.} As a result, for droplets with lower impact velocity, the kinetic energy is lower which leads to smaller spreading diameter compared to droplets with higher velocity [Figure~\ref{fig:velChar}(a)].

For viscoelastic non-Newtonian liquids, elastic energy significantly influences the spreading behavior. As the droplet spreads, the polymer chains within the liquid experience stress and store part of the droplet's kinetic energy as elastic energy. Consequently, both viscous dissipation and elastic restoring forces oppose the spreading motion, thereby restricting the droplet's maximum spread. As a result, \NREV{at a fixed Weber number (with the same nominal diameter and adopted fluid properties)}, the maximum spreading decreases with increasing polymer concentration. However, when the impact velocity is increased, the droplet possesses greater initial kinetic energy, leading to a larger maximum spreading despite the presence of elastic effects [Figure~\ref{fig:velChar}(a)].

The counter-jet height is the other parameter that has been considered. As can be seen in Figure~\ref{fig:velChar}(b), the counter-jet growth is similar for all cases till $\ts \approx 2$. After this, the counter-jet height decreases \NREV{as the impact velocity decreases}, or the solution concentration increases. For the \LANGCUT{0.30} \LANG{0.25}\% solution \REV{hollow} droplet with $U_{0}=2.4$~m/s, \NREV{the counter-jet height approaches a plateau} after $\ts \approx 7$, \NREV{indicating that its upward growth has nearly stopped.} This can be due to the lower kinetic energy available to be transferred to the counter-jet and also \CUT{higher dissipation of energy through higher viscosity} \NEW{the polymer contribution to the} energy \NEW{of the impact: the work done on the stretched chains is stored elastically and dissipated as they relax.}

\CUT{The effect of lower kinetic energy of droplet before the impact is also can be seen in the tip velocity of counter-jets for different solutions [the former appendix figure(a)]. For the droplet impacts in the range of $U_{0}=2.4$--$3.8$~m/s, the counter-jets show similar behavior till $\ts \approx 2$. After this time, the counter-jet velocity differs in each case. Generally, as was observed in Figure~\ref{fig:cjChar}(a), as the polymer concentration increases the maximum counter-jet velocity decreases. The same thing can be seen in for impact velocity decrement [the former appendix figure(a)]. As the droplet impact velocity decreases, the counter-jet tip velocity decreases and the ratio of this reduction is higher than the reduction in the droplet impact velocity, which means at this regime, the viscous and capillary forces suppress inertia. Moreover, the counter-jet tip velocity reduces further by increasing polymer concentration, for instance, in the case of solution 0.30\% impacting at $U_{0}=2.4$~m/s, the counter-jet tip velocity gets close to zero at $\ts \approx 9$.}

\NEW{\NREV{To estimate the maximum spreading diameters reported above, we use a reduced energy balance between the moment just before impact and the moment of maximum spreading.} The droplet arrives with the kinetic energy of its liquid volume and the surface energy of its two interfaces. \NREV{The balance represents the surface energy of the flattened lamella, viscous dissipation in the boundary layer beneath it, energy taken up by polymer deformation and relaxation, and an effective bubble-work term. Residual motion of the jet and liquid sheet is not resolved.} Following Chandra and Avedisian \cite{chandra1991collision} and Pasandideh-Fard \textit{et al.}\ \cite{pasandideh1996capillary}, and extending their balance to the hollow geometry and to the polymer, this gives
\begin{equation}
\left(SD^{*}_{max}\right)^{2}=\frac{\dfrac{\We}{12}+\left(\dfrac{\Dh}{\Deq}\right)^{2}+\left(\dfrac{\Db}{\Deq}\right)^{2}-\dfrac{C_{b}}{12}\We\left(\dfrac{\Db}{\Deq}\right)^{3}}{\NREV{\dfrac{1-\cos\theta}{4}}+\dfrac{\We}{3\sqrt{\Rey_{s}}}+\dfrac{C_{e}\,\eta_{p}\Deq}{6\lambda\gamma}},
\label{eq:sdmax}
\end{equation}
where $\theta$ is \MREV{an effective contact angle for the substrate}, $\Rey_{s}=\rho U_{0}\Deq/\eta_{s}$ is a Reynolds number built with the solvent viscosity, and $C_{e}$ and $C_{b}$ are empirical constants for the polymer contribution and the bubble work. \MREV{For water, the polymer contribution in the denominator is set to zero. For a dense Newtonian droplet ($\Dh=\Deq$, $\Db=0$),} Eq.~(\ref{eq:sdmax}) reduces exactly to the model of Pasandideh-Fard \textit{et al.}\ \cite{pasandideh1996capillary}.}

\begin{figure}[h!]
	\centering
	\includegraphics[width=0.55\textwidth]{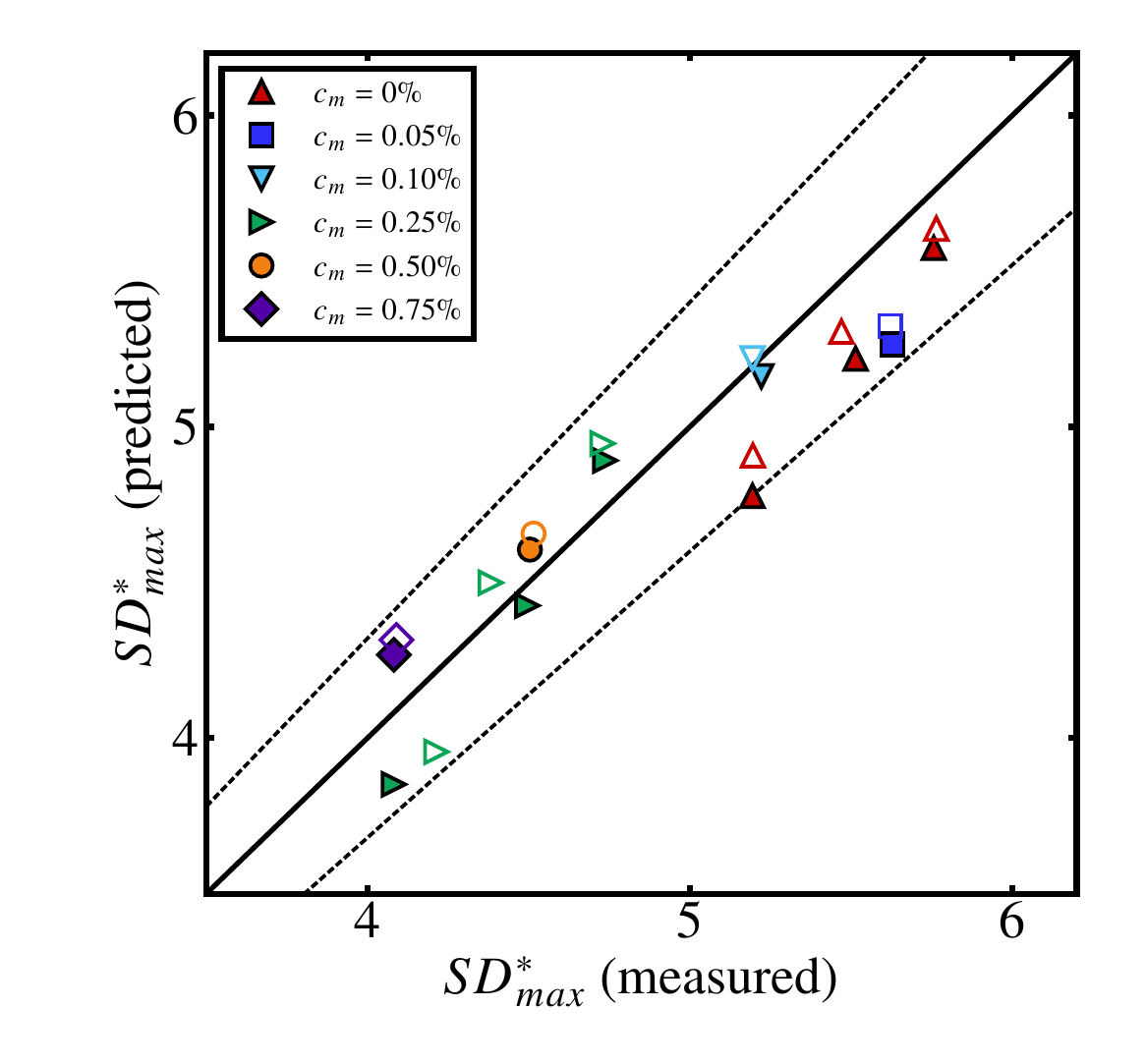}
	\caption{\NEW{Measured against predicted maximum spreading diameter for the dense (filled symbols) and hollow (open symbols) impact conditions, at $U_{0}=3.8$, $3.1$ and $2.4$~m/s. Predictions from Eq.~(\ref{eq:sdmax}) \MREV{\NREV{with $\theta=90^\circ$, and $C_{e}=39.1$ fitted to the seven polymeric dense points}}. The solid line is perfect agreement and the dashed lines mark $\pm8\%$.}}
	\label{fig:sdparity}
\end{figure}

\NEW{Figure~\ref{fig:sdparity} compares Eq.~(\ref{eq:sdmax}) with the measured values. \MREV{\NREV{With the baseline choice $C_{b}=0$}, least-squares fitting of $C_{e}$ to the seven polymeric dense-drop points gives $C_{e}=39.1$. The same coefficient is then applied to the hollow droplets without refitting. Across the 20 plotted dense and hollow conditions at $U_{0}=2.4$, 3.1, and 3.8~m/s, the maximum absolute relative discrepancies are below 8.1\% and 6.3\%, respectively. The calculation uses the mean DoS times and $\eta_{p}=\eta_{0}-\eta_{s}$ from Table~\ref{tab:fluids}.} The derivation and assumptions are given in Appendix~\ref{app:sdmax}.}

\subsection{Regime Map}
\label{sec:regime}

\begin{figure}[htbp]
	\centering
	\includegraphics[width=0.92\textwidth]{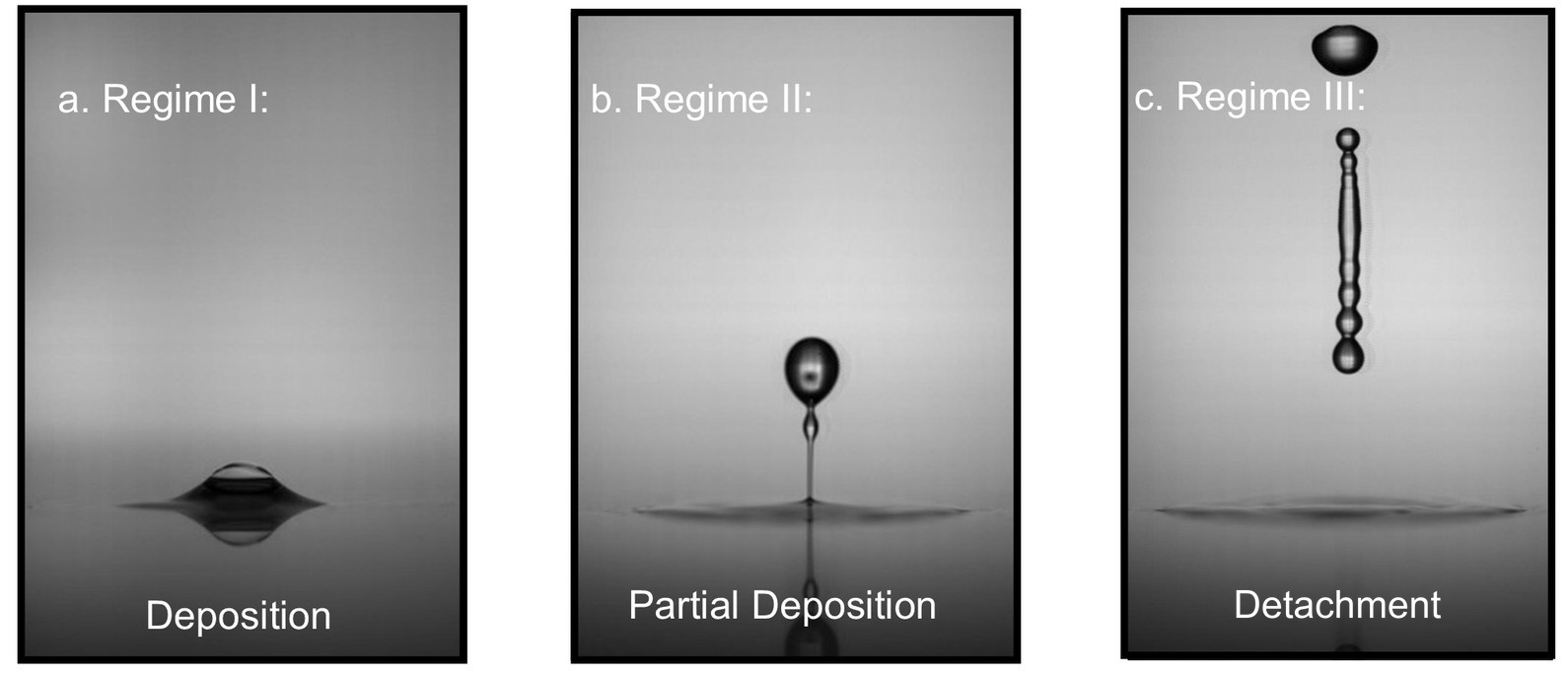}\\[0.6em]
	\includegraphics[width=0.62\textwidth]{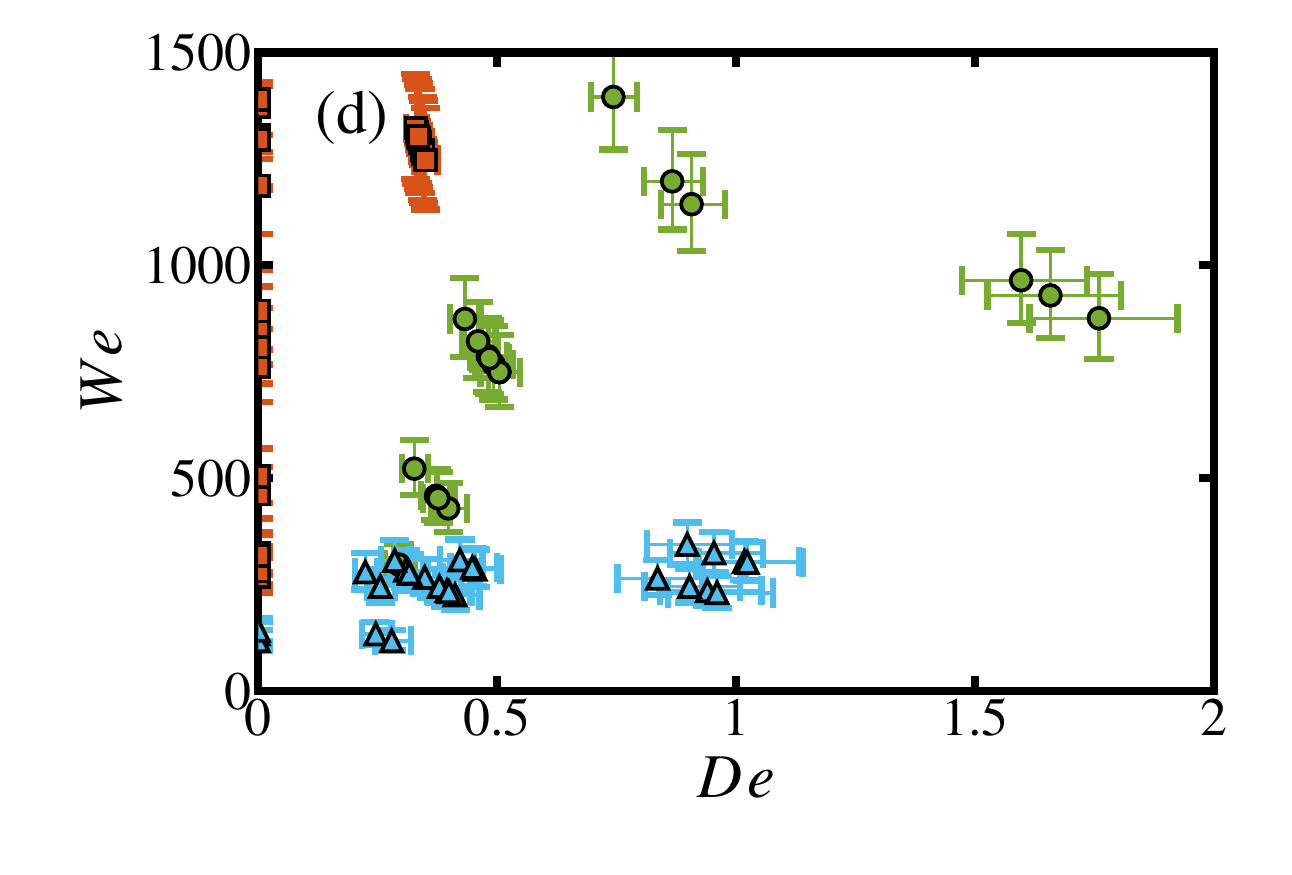}
	\caption{Deposition regimes of hollow droplet impact.
	\NREV{Regimes observed upon hollow droplet impact on the surface:} (a) deposition, (b) partial deposition, and (c) \FIXCUT{rebound} \FIX{detachment} \NEW{regimes\REV{. (d) Observed regimes in $\We$--$\De$ phase space. \MREV{Each symbol represents one experimental impact}, with different symbols denoting the observed impact regime. Blue triangle, green circle, and orange square symbols represent deposition, partial deposition, and \FIXCUT{rebound} \FIX{detachment}, respectively. \MREV{Error bars show propagated diameter and velocity bounds as defined in the text, not standard deviations.}}}}
	\label{fig:regimes}
\end{figure}

Based on the variation in polymer concentration and impact velocity, three distinct regimes are observed upon the impact of the hollow droplets: deposition [Figure~\ref{fig:regimes}(a)], partial deposition [Figure~\ref{fig:regimes}(b)], and \FIXCUT{rebound} \FIX{detachment} [Figure~\ref{fig:regimes}(c)]. Upon impact, the droplet interacts with the surface, and depending on the balance among inertial, capillary, and elastic forces, different outcomes emerge. In cases with low inertia, the deposition regime is observed, where the counter-jet lacks sufficient energy to rupture the bubble shell and remains trapped within it. At higher inertia, the counter-jet possesses enough kinetic energy to rise, \REV{rupture} the shell and \MREV{detach the counter-jet from the liquid remaining on the substrate}, resulting in the \FIXCUT{rebound} \FIX{detachment} regime. \NREV{This regime occurs in both polymeric and Newtonian droplets.} The remaining regime, partial deposition, \MREV{is observed here in polymer solutions}, depending on the strength of the elastic forces. In this regime, while the counter-jet \NEW{\REV{ruptures}} the bubble shell, the presence of the elastic forces leads to formation of BOAS structures, preventing the \FIXCUT{rebound} \FIX{detachment} of the \MREV{counter-jet during the observation interval}.

Accordingly, the observed regimes are mapped onto the $\We$--$\De$ phase space [(Figure~\ref{fig:regimes}(d)]. \MREV{The 70 observations comprise 27 deposition, 18 partial-deposition, and 25 counter-jet-detachment events.} \NREV{For Newtonian droplets, deposition occurs at low} $\We$ numbers, \NREV{while increasing $\We$ transitions the system} into the \FIXCUT{rebound} \FIX{detachment} regime. In polymeric droplets, the deposition regime shifts to higher $\We$ values due to the added resistance from elastic stresses. When $\De$ is small and $\We$ is sufficiently high, the inertial forces remain dominant, resulting in formation of the \FIXCUT{rebound} \FIX{detachment} regime. Finally, the partial deposition regime emerges in an intermediate region where $\We$ and $\De$ compete.

\clearpage
\section{Conclusion}
\label{sec:conclusion}

This experimental study has delineated the complex dynamics of hollow non-Newtonian droplet impact on a solid surface, revealing phenomena distinct from those observed in dense droplet impacts. \NREV{The entrapped bubble promotes a central counter-jet and alters the subsequent deposition dynamics.}

The investigation into polymeric solutions has yielded several key insights:

\textbf{Spreading Suppression:} \NREV{Increasing polymer concentration reduces the maximum spreading diameter and the counter-jet height in the comparisons presented, consistent with energy taken up by polymer deformation and relaxation.}

\textbf{Altered Filament Dynamics:} Viscoelasticity fundamentally changes the thinning and breakup process of the counter-jet. Instead of rapid pinch-off, we observe the formation of \MREV{persistent} filaments and beads-on-a-string structures, a direct manifestation of the interplay between elastic stresses resisting extension and capillary pressure promoting breakup. \MREV{Exponential fits give apparent impact-filament relaxation times of $1.5$ to $10.7$~ms, consistent with an elasto-capillary interpretation over the fitted intervals. Their dependence on fit interval, geometry, and deformation history limits their identification as material relaxation times.}

\textbf{Delayed Dynamics:} Higher polymer concentrations delay both bubble rupture and counter-jet detachment from the surface, extending the lifetime of the filament and modifying the splash outcome.

\NEW{\textbf{Analytical Model for Maximum Spreading:} An energy balance between impact and maximum spreading, written for a hollow droplet and including the energy taken up by the stretched polymer chains, gives a closed-form expression for the maximum spreading diameter, Eq.~(\ref{eq:sdmax}). \MREV{With $C_e=39.1$ fitted to the polymeric dense-drop data, discrepancies are below 8.1\% for the dense and 6.3\% for the hollow conditions plotted,} and it accounts for the fall of the spreading diameter with concentration reported above.}

\textbf{Regime Classification:} The interplay of inertial, capillary, and elastic forces leads to three distinct outcomes---deposition, partial deposition, and \FIXCUT{rebound} \FIX{detachment}---which we successfully map onto a Weber--Deborah number phase space.

This work provides a foundational understanding of the parameters controlling hollow viscoelastic droplet impact. The insights gained, particularly the mapping of deposition regimes, are valuable for optimizing industrial processes such as spray coating with structured fluids or inkjet printing of complex formulations. The \NEW{\MREV{analytical expression estimates the maximum spreading diameter from the fluid properties, impact conditions, and a coefficient calibrated on dense-drop measurements. Its agreement with the hollow-drop data without refitting supports its use within the tested conditions.} The} rich phenomena observed, especially the \CUT{electrocapillary} \NEW{elasto-capillary} bead formation, invite further numerical modeling to quantitatively predict the dynamics explored here experimentally. \NREV{Although the authors have previously modeled Newtonian hollow droplet impact and bubble collapse~\cite{nasiri2024experimental,nasiri2021hollow,nasiri2023flattening,moezzirafie2018investigation}, and numerical studies of dense polymeric droplet impact exist, a numerical investigation of the coupled processes examined here remains a subject for future work.} Future work will focus on developing a robust numerical framework to complement these experimental findings and provide a more predictive capability for this complex multiphase flow.

\begin{acknowledgments}
We gratefully acknowledge funding support from 3M Company.
\end{acknowledgments}

\section*{Declaration of interests}
The authors declare that they have no known competing financial interests or personal relationships that could have appeared to influence the work reported in this paper.

\clearpage
\appendix
\renewcommand{\thefigure}{A\arabic{figure}}
\setcounter{figure}{0}
\section{\ed{Supplementary snapshots}}
\label{app:snapshots}

\begin{figure}[htbp]
	\centering
	\includegraphics[width=0.95\textwidth]{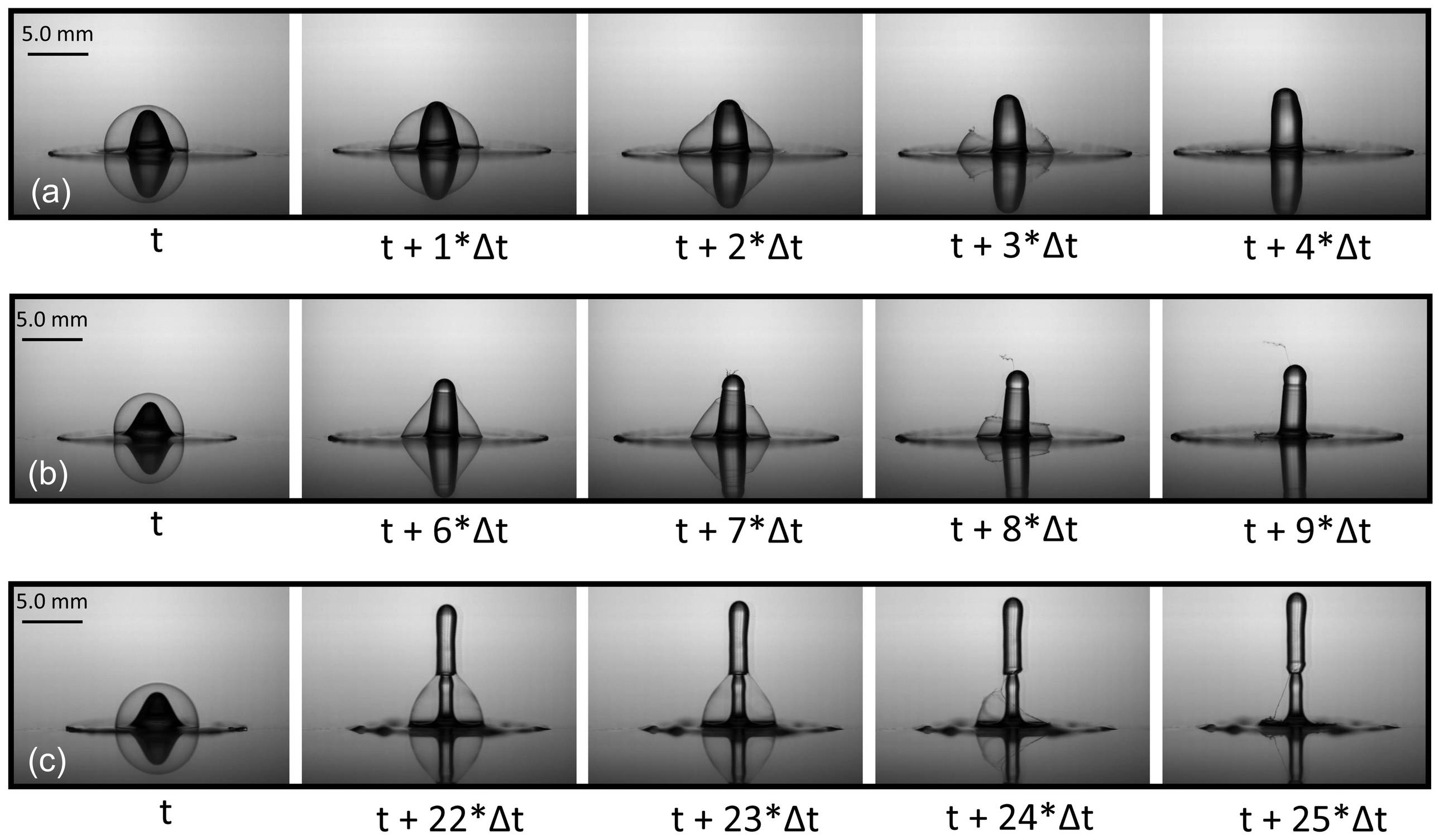}
	\caption{Selected snapshots of bubble rupture after a hollow droplet impact on an aluminum surface at $U_{0}=3.8$~m/s, $\We\approx880$, $\Oh\approx0.0016$.
	(a) A water droplet without polymer. (b) A water droplet with 0.10\% polymer. (c) A water \NEW{droplet} with 0.50\% polymer.}
	\label{fig:Arupture}
\end{figure}

\MREV{Figure~\ref{fig:Arupture} shows selected stages of bubble rupture. The first-to-last displayed spans are $4\Delta t$, $9\Delta t$, and $25\Delta t$ for water, 0.10\% PEO, and 0.50\% PEO, respectively, with $\Delta t=0.0002$~s; these correspond to 0.8, 1.8, and 5.0~ms. These spans illustrate the evolution of rupture but do not establish its duration from a common jet--shell contact event. Slower jet motion and polymeric stresses in the liquid shell may contribute to the delayed rupture at higher concentrations.}

\begin{figure}[htbp]
	\centering
	\includegraphics[width=0.95\textwidth]{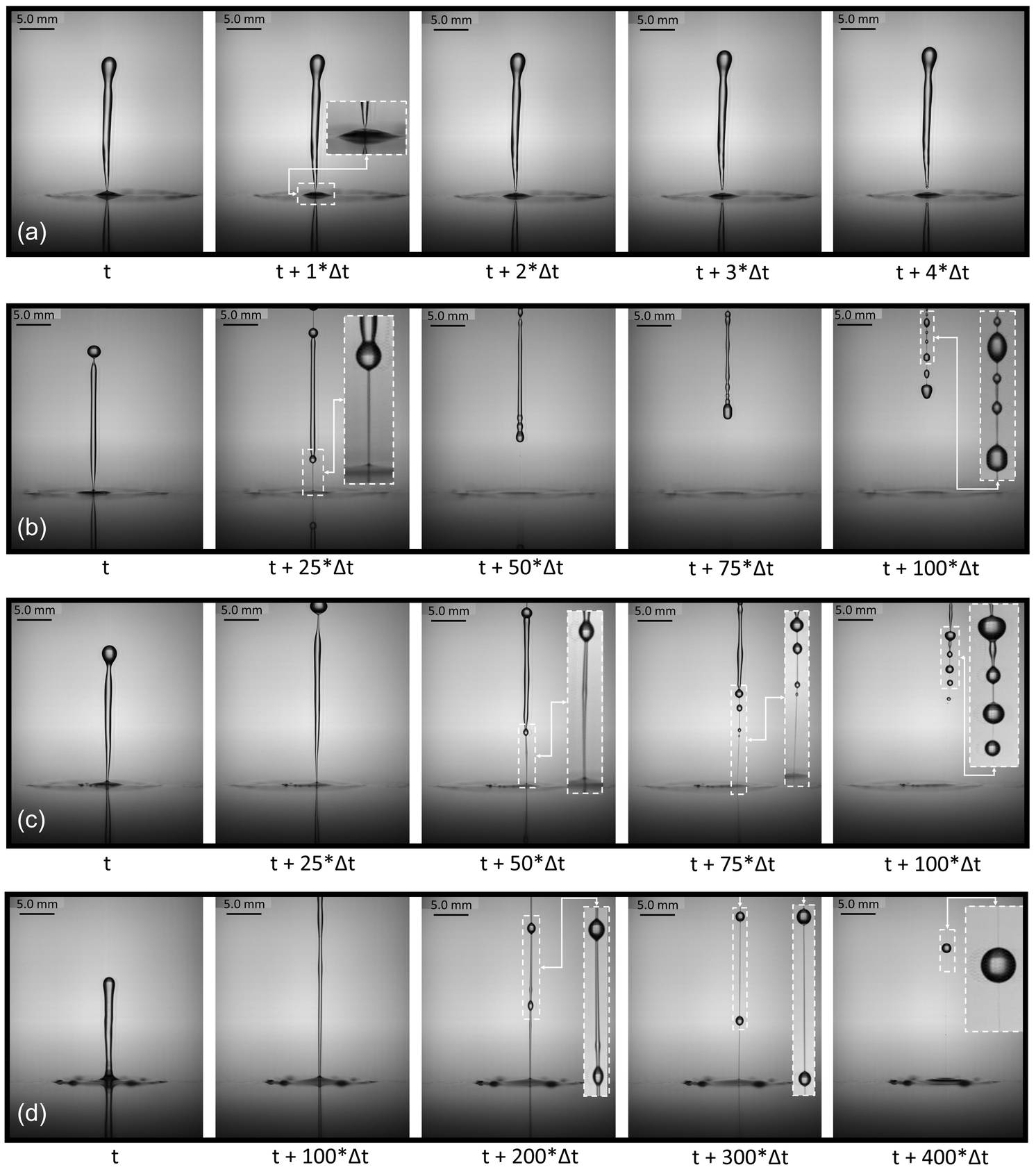}
	\caption{Selected snapshots of counter-jet breakup from the surface after a hollow droplet impact on an aluminum surface at $U_{0}=3.8$~m/s, $\We\approx880$, $\Oh\approx0.0016$.
	(a) A water droplet without polymer. (b) A water droplet with 0.05\% polymer. (c) A water \NEW{droplet} with 0.10\% polymer. (d) A water \NEW{droplet} with 0.50\% polymer.}
	\label{fig:Abreakup}
\end{figure}

Figure~\ref{fig:Abreakup} shows selected snapshots of \NEW{counter-jet breakup} after a hollow droplet impact on an aluminum surface. \MREV{The first-to-last displayed spans are $4\Delta t$, $100\Delta t$, $100\Delta t$, and $400\Delta t$ for water, 0.05\%, 0.10\%, and 0.50\% PEO, respectively. With $\Delta t=0.0002$~s, these are 0.8, 20, 20, and 80~ms. They are snapshot spans rather than breakup times measured from a common initial state.} A distinct behavior visible in Figure~\ref{fig:Abreakup} is the emergence of beads-on-a-string (BOAS) patterns. In this configuration, nearly cylindrical threads link a sequence of droplet-like beads, creating a pearl-necklace-type appearance. \MREV{This morphology is consistent with the capillary--elastic competition discussed in Section~\ref{sec:polymeric}.}

\CUT{the former appendix figure shows snapshots of water hollow droplet impacting on the surface at different velocities.}

\CUT{the former appendix figure displays the evolution of counter-jet for Newtonian and different concentrations droplets. For the Newtonian droplet, the filament thickness exhibits a sharp decay until it reaches zero, indicating the detachment of the droplet from the surface. In contrast, the evolution of filament thickness in polymer solutions follows a distinct pattern.}

\CUT{the former appendix figure shows snapshots of polymer solution hollow droplet with 0.3\% concentration impacting on the surface at different velocities.}

\clearpage
\section{\tb{Derivation of the spreading-diameter model}}
\label{app:sdmax}

\NEW{This appendix derives the spreading-diameter model, Eq.~(\ref{eq:sdmax}), \FIXCUT{term by term, following} \FIX{using energy balance in line with} Refs.~\onlinecite{chandra1991collision,pasandideh1996capillary}.}

\NEW{Before impact the droplet carries the kinetic energy of its liquid volume and the surface energy of its two liquid--gas interfaces, the outer shell and the bubble,
\begin{equation}
E_{0} = E_{k}+E_{s,0}=\frac{1}{2}\rho\left(\frac{\pi}{6}\Deq^{3}\right)U_{0}^{2} + \pi\gamma\left(\Dh^{2}+\Db^{2}\right),
\label{eq:E0}
\end{equation}
where the liquid volume is written with the equivalent diameter of Eq.~(\ref{eq:Deq}) and the mass and viscosity of the entrapped air are neglected. Between impact and maximum spreading this energy is distributed over four sinks,
\begin{equation}
E_{0} = E_{s,f} + E_{v} + E_{e} + E_{b},
\label{eq:balance}
\end{equation}
with the macroscopic kinetic energy at maximum spreading taken as zero \cite{pasandideh1996capillary}. \MREV{Gravity is neglected during the early spreading stage:} the Froude number $U_{0}^{2}/(g\Deq)$ \NREV{exceeds $130$ for the maximum-spreading comparisons at $U_0=2.4$--$3.8$~m/s}.}

\NEW{The first sink is the surface energy of the liquid at maximum spreading, when the liquid is a thin disc of diameter $SD_{max}$. Its free surface, of area $(\pi/4)SD_{max}^{2}$, carries the energy density $\gamma$. Its wetted base, of the same area, replaces a solid--gas interface by a solid--liquid one, which costs $\gamma_{sl}-\gamma_{sv}=-\gamma\cos\theta$ per unit area by Young's equation, \MREV{with $\theta$ treated here as an effective model contact angle.} Adding the two contributions and neglecting the thin rim and the transient central hole gives \cite{pasandideh1996capillary,ukiwe2005maximum}
\begin{equation}
\NREV{E_{s,f} = \frac{\pi}{4}\,SD_{max}^{2}\,\gamma\left(1-\cos\theta\right).}
\label{eq:Esf}
\end{equation}}

\NEW{The second sink is the viscous dissipation in the boundary layer under the spreading lamella. Following Ref.~\onlinecite{pasandideh1996capillary}, the dissipation rate per unit volume is $\Phi\approx\eta_{s}\left(U_{0}/\delta\right)^{2}$, and it acts over the boundary-layer volume $(\pi/4)SD_{max}^{2}\,\delta$ for the spreading time $t_{c}=8\Deq/(3U_{0})$, with $\delta=2\Deq/\sqrt{\Rey_{s}}$ the boundary-layer thickness of an axisymmetric stagnation flow. Combining these, and using $\eta_{s}U_{0}\sqrt{\Rey_{s}}=\rho U_{0}^{2}\Deq/\sqrt{\Rey_{s}}$,
\begin{equation}
E_{v} \approx \Phi\,\frac{\pi}{4}\,SD_{max}^{2}\,\delta\,t_{c} = \frac{\pi}{3}\,\frac{\rho U_{0}^{2}\,\Deq\,SD_{max}^{2}}{\sqrt{\Rey_{s}}},
\qquad \Rey_{s}=\frac{\rho U_{0}\Deq}{\eta_{s}}.
\label{eq:Ev}
\end{equation}

\MREV{Using the solvent viscosity in this boundary-layer estimate is a modeling approximation. The available shear-viscosity measurements do not independently establish the high-rate viscosity during impact.}}

\NEW{The third sink is the energy taken up by the polymer as the lamella stretches the chains \cite{bergeron2000controlling}. \MREV{Motivated by the single-mode Maxwell (Oldroyd-B) relation $G=\eta_{p}/\lambda$~\cite{entov1997effect,anna2001elasto}, we use this ratio as an effective polymer-stress scale and $SD_{max}/\Deq$ as a characteristic stretching measure. The resulting energy closure over the liquid volume is}
\begin{equation}
E_{e} = C_{e}\,\frac{\eta_{p}}{\lambda}\left(\frac{SD_{max}}{\Deq}\right)^{2}\frac{\pi}{6}\Deq^{3},
\label{eq:Ee}
\end{equation}}

\NEW{The last sink is the work spent compressing and rupturing the bubble, a feature specific to the hollow droplets \cite{gulyaev2013hollow,nasiri2023flattening}. The pressure acting on the bubble in the early stage of impact is of the order of the stagnation pressure $\frac{1}{2}\rho U_{0}^{2}$, so this work is written as a fraction $C_{b}$ of the stagnation pressure acting over the bubble volume,
\begin{equation}
E_{b} = C_{b}\,\frac{1}{2}\rho U_{0}^{2}\,\frac{\pi}{6}\Db^{3} = C_{b}\,\frac{\pi}{12}\,\rho U_{0}^{2}\Db^{3}.
\label{eq:Eb}
\end{equation}}

\MREV{The calculations use $C_{b}=0$ as a baseline assumption, not as a measurement of zero bubble work. Similar maximum spreading of dense and hollow droplets does not uniquely determine $C_b$, because their initial interfacial energies differ. The polymer coefficient $C_e$ is calibrated to the dense-drop data, and the contact angle and solvent-viscosity closure are additional approximations. Residual jet motion and bubble dynamics are not resolved by this reduced energy balance, which limits extrapolation beyond the tested conditions.}

\clearpage
\bibliography{references_v2}

\end{document}